\documentclass[print]{aa} 
\usepackage{savesym}%
\usepackage{graphicx}%
\usepackage{hyperref}%
\usepackage{lineno}%
\usepackage{mathtools}%
\usepackage{amsmath}%
\usepackage{xspace}%
\usepackage{units}%
\usepackage{relsize}%
\usepackage{tikz}%
\usepackage{amssymb,amsfonts,amsmath,amstext,amsgen,amsopn,amsxtra,indentfirst,times}%
\usetikzlibrary{positioning}%
\usetikzlibrary{shapes,arrows}%
\usepackage{nicefrac}%
\usepackage{array}

\DeclareFontFamily{U}{mathb}{\hyphenchar\font45}
\DeclareFontShape{U}{mathb}{m}{n}{
      <5> <6> <7> <8> <9> <10> gen * mathb
      <10.95> mathb10 <12> <14.4> <17.28> <20.74> <24.88> mathb12
      }{}
\DeclareSymbolFont{mathb}{U}{mathb}{m}{n}
\DeclareMathSymbol{\Earth}{3}{mathb}{"43}

\usepackage{tabularx}  
\usepackage{ragged2e}  
\newcolumntype{Y}{>{\RaggedRight\arraybackslash}X} 
\usepackage{booktabs}  

\newcommand{\coco}{\color{black}}
\newcommand{\Pbatm}{$p_{\rm batm}$\xspace}

\newcommand{\mice}{$m_{\rm water}$\xspace}

\newcommand{\RE}{R$_{\rm \Earth}$\xspace}
\newcommand{\ME}{M$_{\rm \Earth}$\xspace}

\newcommand{\Zenv}{$Z_{\rm env}$\xspace}
\newcommand{\rc}{$r_{\rm core}$\xspace}
\newcommand{\rsolid}{$r_{\rm mantle}$\xspace}
\newcommand{\menv}{$m_{\rm env}$\xspace}
\newcommand{\prior}{{\rm p}({\bf m})\xspace}
\newcommand{\m}{{\bf m}\xspace}
\newcommand{\dat}{{\bf d}\xspace}
\newcommand{\post}{{\rm p}({\bf m}|{\bf d})\xspace}
\newcommand{\fesi}{{\rm Fe}/{\rm Si}_{\rm bulk}\xspace}
\newcommand{\mgsi}{{\rm Mg}/{\rm Si}_{\rm bulk}\xspace}
\newcommand{\fesistar}{{\rm Fe}/{\rm Si}_{\rm star}\xspace}
\newcommand{\mgsistar}{{\rm Mg}/{\rm Si}_{\rm star}\xspace}
\newcommand{\fesima}{{\rm Fe}/{\rm Si}_{\rm mantle}\xspace}
\newcommand{\mgsima}{{\rm Mg}/{\rm Si}_{\rm mantle}\xspace}

\begin{document}

   \title{A generalized bayesian inference method for constraining the interiors of super Earths and sub-Neptunes}
   \author{Caroline Dorn\inst{1}
              \and Julia Venturini\inst{1}
              \and Amir Khan\inst{2}
              \and Kevin Heng\inst{1}
              \and Yann Alibert\inst{1}
              \and Ravit Helled\inst{3}
              \and Attilio Rivoldini\inst{4}
              \and Willy Benz\inst{1}}
 \authorrunning{Dorn et al.}
\titlerunning{A Generalized Bayesian Inference Method for Constraining Planet Interiors} 

          \institute{Physics Institute, University of Bern, Sidlerstrasse 5, CH-3012, Bern, Switzerland\\
              \email{caroline.dorn@space.unibe.ch}
         \and
             Institute of Geophysics, ETH Z\"urich, Sonneggstrasse 5, 8092 Z\"urich\\
             \and
             Department of Geosciences, Raymond \& Beverly Sackler Faculty of Exact Sciences, Tel Aviv             University, Tel Aviv, 69978, Israel \\
             \and Royal Observatory of Belgium, Earth Rotation and Space Geodesy, B-1180 Bruxelles, Belgium\\
             }
             

 
\abstract{}
   {We aim to present {\coco a} generalized Bayesian inference method for constraining  interiors of super Earths and sub-Neptunes. Our methodology succeeds in  quantifying the degeneracy and correlation of  structural parameters for high dimensional parameter spaces. Specifically, we identify what constraints can be placed on {\coco composition and thickness of core, mantle, ice, ocean, and atmospheric layers} given observations of mass, radius, and bulk refractory abundance constraints (Fe, Mg, Si) from observations of the host star's photospheric composition. 
    }
   {We employed a full probabilistic Bayesian inference analysis that formally accounts for observational and model uncertainties. Using a Markov chain Monte Carlo technique, we computed joint and marginal posterior probability distributions for all structural parameters of interest. We included state-of-the-art structural models {\coco based on} self-consistent thermodynamics of core, mantle, high-pressure ice, and liquid water. Furthermore, we tested and compared two different atmospheric models that are tailored for modeling thick and thin atmospheres, respectively.}
   {First, we validate our method against Neptune. Second, we apply it to synthetic exoplanets of fixed mass and determine the effect on interior structure and composition when (1)  radius, (2) atmospheric model, (3) data uncertainties, (4)  semi-major axes, (5) atmospheric composition (i.e., a priori assumption of enriched envelopes versus pure H/He envelopes), and {\coco (6) prior distributions} are varied.}
   {Our main conclusions are: (1) Given available data, the range of possible interior structures  is large; quantification of the degeneracy of possible interiors is therefore indispensable for meaningful planet characterization.
(2) Our method predicts models that agree with independent estimates of Neptune's interior.
(3) Increasing the precision in mass and radius leads to much improved constraints on ice mass fraction, size of rocky interior, but little improvement in the composition of the gas layer, whereas an increase in the precision of stellar abundances enables to better constrain mantle composition and relative core size.
(4) For thick atmospheres, the choice of atmospheric model can have significant influence on interior predictions, {\coco including} the rocky and icy interior. The preferred atmospheric model is determined by envelope mass.

 This study provides a methodology for rigorously analyzing general interior structures of exoplanets which may help to understand how exoplanet interior types are distributed among star systems. This study is relevant in the interpretation of future data from missions such as TESS, CHEOPS, and PLATO.}


   \maketitle

\section{Introduction}
\sloppy
The characterization of planet interiors is one of the main foci of current exoplanetary science. 
For the characterization of super Earths and sub-Neptunes, we mostly rely on mass and radius measurements.   Direct measurements of atmospheres are, thus far, mostly limited to transiting hot Jupiters and {\coco a} few Sub-Neptunes \citep{iyer}, with the exception of super Earth 55 Cnc E \citep{Tsiaras, demory}. For interior characterization, common practice is the use of mass-radius-plots where
mass and radius of exoplanets are compared to synthetically computed interior models \citep[e.g.,][]{sotin07, seager2007, fortney, dressing, howe}. {\coco However}, it is difficult to know (1)  how well one interior model compares with the generally large number of other possible interior scenarios that also fit data and (2) which structural parameters can actually be constrained by the observations.
Thus, this approach fails to address {\coco the degeneracy problem} that is, that different interior models can have identical mass and radius. In order to draw meaningful conclusions about an exoplanet's interior it  is  therefore necessary to account for this inherent degeneracy {\citep[e.g.,][]{rogers2010, schmitt2014, carter, weiss, dorn}}. 

The Bayesian analysis of \citet{rogers2010} to exoplanets of three to four parameters was generalized for purely rocky exoplanets by \citet{dorn}. Here, we extend the full probabilistic analysis of \cite{dorn} to more general interior structures by including volatile elements in form of icy layers, oceans, and atmospheres. {\coco The previous work of \citet[][]{rogers2010} uses a grid search method which calls for strong a priori assumptions on  structure and composition of exoplanets to significantly reduce the parameter space. However, the number of parameters that affect mass and radius is large (e.g., it comprises composition and size of core, mantle, ice layers, and gas, as well as internal energy). Here, we present a generalized Bayesian inference scheme that incorporates the following aspects}:
\begin{itemize}  \itemsep0pt
\item Our method is applicable to a wide range of planet-types, including rocky super Earths and sub-Neptunes.

\item We employ a full probabilistic Bayesian inference analysis using a Markov chain Monte Carlo (McMC) technique to constrain core size, mantle thickness and composition, mass of water-ice, and key characteristics of the atmosphere (e.g., mass, intrinsic luminosity, composition).

\item We test two different atmospheric models, tailored to thick and thin atmospheres, that account for enrichments in elements heavier than H and He. 

\item We employ state-of-the-art  modeling to compute interior structure based on self-consistent thermodynamics for a pure iron core, a silicate mantle, high-pressure ice, water ocean, and atmosphere (to some extent). 

 \item {\coco Compared to previous work of \citet[][]{rogers2010}, our scheme can also be used for high dimensional parameter spaces.}
\end{itemize}

 Besides mass and radius estimates, additional constraints are crucial to reduce model degeneracy  \citep[e.g.,][]{dorn, grasset09}. 
\citet{dorn} demonstrate that the use of relative bulk abundance constraints of Fe/Si and Mg/Si taken from the host star (henceforth referred to as abundance constraints) leads to much improved constraints on core size and mantle composition in the case of purely rocky exoplanets. The validity of a direct correlation between  stellar and planetary relative bulk abundances is suggested by observational solar system studies  and planet formation models \citep{carter, lodders03,drake,mcdono, bond,elser,johnson,thiabaud}. Here, we also assume solar bulk abundance constraints based on spectroscopic measurements \citep{lodders03}. 

Our generalized interior structure model is based on previous studies of mass-radius relations.
Generally, H$_2$O in liquid and high-pressure ice form \citep[e.g.,][]{valencia07a, seager2007}, and H$_{2}$-He atmospheres \citep[e.g.,][]{rogers2011, fortney} are considered.  Although it would not be surprising if the compositional diversity of ices and atmospheres exceeds the one found in the solar system \citep[e.g.,][]{newsom}, the few observational data on exoplanets limit us to relatively simple planetary interior models. 

The structural parameters that we {\coco investigate} include: (1) internal energy, mass, and composition of the gas layer, (2) mass and temperature of the ice layer, (3) mantle size and composition, and (4) core size. For present purposes, we assume a general planetary structure consisting of a pure iron core, a silicate mantle, a water ice layer and an atmosphere.
To compute the resultant density profile for the purpose of estimating mass and radius, we follow \citet[][]{dorn} and assume hydrostatic equilibrium coupled with a thermodynamic approach based on Gibbs free-energy minimization and
Equation-of-State (EoS) modeling. 

In this study, we wish to quantify the influence of the following parameters on predicted interior structure and composition: (1) planet radius, (2) data uncertainty (e.g., mass, radius, bulk abundances), (3)  semi-major axis, (4) atmospheric model, (5) atmospheric composition (i.e., a priori assumption of enriched envelopes versus pure H/He envelopes), and {\coco (6) prior distributions.}
{\coco
In a companion paper \citep{dornA}, we present results on the application of our proposed method to six exoplanets (\mbox{HD 219134b}, Kepler-10b, Kepler-93b, CoRoT-7b, 55 Cnc e, and \mbox{HD 97658b}) for which spectroscopic measurement of their host star's photospheres are available \citep{hinkel}.}

The outine of this study is as follows: we describe the iterative inference scheme (Section \ref{inversion}), model parameters (Section \ref{parametrization}), data (Section \ref{data}), and the forward model (Section \ref{model}). In Section 3, we validate our method against Neptune and present results for different synthetic planet cases. In Sections \ref{Discussion} and \ref{Conclusions}, we discuss results and conclude.

\section{Methodology}
\label{Methodology}

\subsection{Bayesian inference}
\label{inversion}

We employ a Bayesian method to compute the posterior probability density function (pdf) for each model parameter \m from data \dat and prior information. According to Bayes' theorem, the posterior distribution $\post$ for a fixed model parameterization, conditional on data, is proportional to prior information $\prior$ on model parameters and the likelihood function ${\rm L}({\bf m}|{\bf d})$, which can be interpreted in probabilistic terms as a measure of how well a given model fits data:

 \begin{equation}
\label{post}
\post \propto \prior {\rm L}({\bf m}|{\bf d}),
\end{equation}
and
\begin{equation}
\begin{aligned}
\label{like}
{\rm L}({\bf m}|{\bf d}) ={} & \frac{1}{(2\pi)^{N/2} (\prod_{i=1}^{N}\sigma_i^2)^{1/2}} \\
& {\rm exp} \left(- \frac{1}{2} \sum_{i=1}^{N}\frac{(g_i({\bf m})-{\bf d_{\rm i}})^{2} }{\sigma_i^2}\right),
\end{aligned}
\end{equation}
where $N$ is the total number of data points, and $\sigma_i$ is the estimated error on the i{\it th} datum.
  In practice, the posterior distribution can not be derived analytically; instead we employ McMC simulation that samples the prior parameter space and evaluates the distance of the response of each candidate model to data. {\coco Finally}, we use the Metropolis-Hastings algorithm to efficiently explore the posterior distribution. 
  
  Briefly, the inference strategy works as follows. An initial starting model is drawn randomly from the prior distribution. The posterior
density of this model is calculated {\coco using} Eq. \ref{post}--\ref{like}.
A new
(candidate) model is subsequently created from a proposal distribution
that is centered around the current model. Moving from the current to the new model is accepted with a probability that depends on their likelihood ratio \citep{MooT}. The method works iteratively and the samples that are generated with this approach are distributed
according to the posterior distribution. We refer to \citet[][]{dorn} for more details.

The large number of models needed for the analysis requires very efficient computations. Presently, generating models of the internal structure of a planet takes on average 40 -- 90 seconds of CPU time on a four quad-core AMD Opteron 8380 CPU node and 32 GB of RAM. {\coco Ten independent McMC chains were run. Burn-in periods (i.e., number of samples until stationary distribution has been reached) last on average some hundred samples. Convergence is reached when the effective length (actual length divided by the autocorrelation length) is large (>1000). In all, we analyzed some $10^5$ models.}

\subsection{Model parameterization}
\label{parametrization}
\begin{figure}[ht]
a) model I\\
\begin{center}
\vspace{-1cm} \mbox{ }
 \includegraphics[width = .33\textwidth]{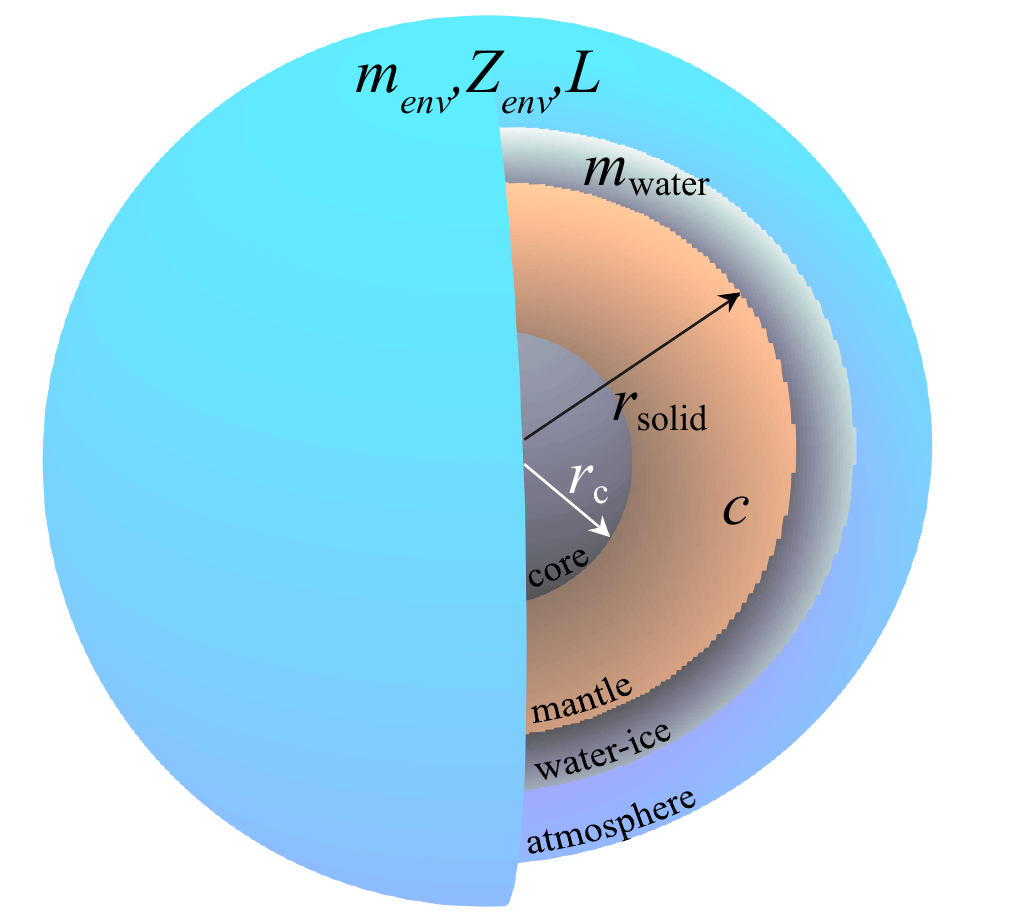} \\
 \end{center}
 b) model II \\
 \begin{center}
 \vspace{-1cm} \mbox{ }
  \includegraphics[width = .33\textwidth]{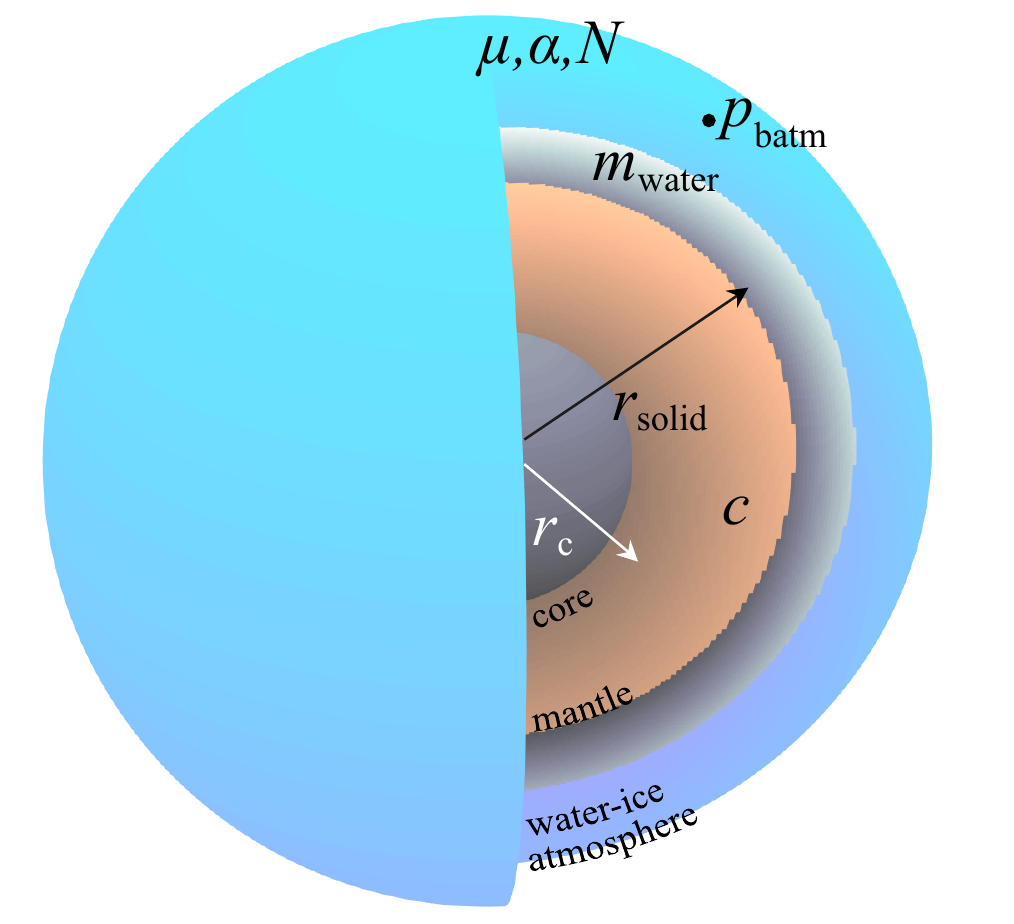}\\
 \end{center}
 \caption{Illustration of model parameterization. (a) Model I parameters are core radius \rc, mantle composition $c$ comprising the oxides Na$_2$O--CaO--FeO--MgO--Al$_2$O$_3$--SiO$_2$, mantle radius \rsolid, mass of water \mice, mass of envelope \menv, envelope Luminosity $L$, and envelope metallicity \Zenv. (b) Model II parameters are as for a), {\coco with} atmosphere parameterized by pressure at the bottom of the atmosphere \Pbatm,
number of scale-heights of opaque layers $N$, 
mean molecular weight $\mu$, and a
temperature-related parameter $\alpha$. See Section \ref{parametrization} and table \ref{momo} for more details. \label{Illustration}}
\end{figure}

Our exoplanet interior model consists of a layered sphere with an iron core surrounded by a silicate mantle, a water layer, and an atmosphere as illustrated in Figure \ref{Illustration}. We distinguish between two different  atmospheric models: a radiative transfer model (model I) and a pressure scale-height model (model II). These models are discussed further in Section \ref{atmomodel}. The key characteristics of both models are parameterized in table \ref{momo}.

\begin{table}[ht]
\caption{Summary of model parameters \m. \Zenv (model I) is defined as the envelope mass fraction of elements heavier than H and He (here C and O).\label{momo}}
\begin{center}
\begin{tabularx}{0.49\textwidth}{@{} l>{\setlength{\baselineskip}{0.8\baselineskip}}Y cl@{}} 
\hline\noalign{\smallskip}
parameter &description& model \\
\noalign{\smallskip}
\hline\noalign{\smallskip}
\rc &core radius & I, II\\
$\fesima$&mantle Fe/Si & I, II\\
$\mgsima$ &mantle  Mg/Si&I, II \\
\rsolid &mantle radius & I, II\\
\mice &mass of water & I, II\\
 \menv& mass of envelope& I\\
 $L$ &envelope Luminosity& I\\ 
 \Zenv & envelope metallicity&I \\
 \Pbatm& pressure at bottom of atmosphere& II\\
 $N$ &number of scale-heights of opaque layers &II\\
$\mu$ &mean molecular weight &II \\
$\alpha$&temperature-related parameter  & II\\
\hline
\end{tabularx} 
\end{center}
\end{table}

\subsection{Data}
\label{data}
The data \dat that we rely on are listed in Table \ref{dada}.

\begin{table}[ht]
\caption{Summary of data \dat. We do not account for uncertainty in those parameters that are labeled as `fixed'.\label{dada}}
\begin{center}
\begin{tabularx}{0.45\textwidth}{@{} l >{\setlength{\baselineskip}{0.8\baselineskip}}Y l @{}} 
\hline\noalign{\smallskip}
parameter & description & comment \\
\noalign{\smallskip}
\hline\noalign{\smallskip}
$M$& planetary mass & \\
$R$& planetary radius  &\\ 
$\fesi$& bulk planetary ratio Fe/Si& \\ 
$\mgsi$& bulk planetary ratio Mg/Si& \\ 
$c_{\rm minor}$ & mantle composition of minor elements: CaO, Al$_2$O$_3$, Na$_2$O& fixed \\ 
$a$&semi-major axis & fixed \\
$R_{\rm star}$ & stellar radius &fixed \\
$T_{\rm star}$& stellar effective temperature &fixed  \\
\hline
\end{tabularx} 
\end{center}
\end{table}

$\fesi$ is {\coco the} mass ratio between the mass of iron to silicate for the entire planet (core and mantle), whereas $\fesima$ is only that which is contained in the mantle. Since all magnesium and silicate are in the mantle, $\mgsi$ equals their mass ratio for the mantle $\mgsima$. {\coco We use the stellar abundances ($\fesistar$ and $\mgsistar$) as a proxy for $\fesi$ and $\mgsi$}. Similarly, we fix the absolute abundance of minor refractory elements (Na, Ca, and Al) in the mantle  $c_{\rm minor}$ to stellar values. {\coco Here, we consider solar estimates for Fe/Si and Mg/Si and associated uncertainties, as well as Na$_2$O, CaO, and Al$_2$O$_3$ using the values of \citep{lodders2009}}.
Stellar radius, and stellar effective temperature are also fixed parameters. {\coco Because uncertainty on stellar radius is generally small compared to uncertainties on planet radius, we neglect possible correlations between both and fix stellar radius.}

\subsection{Structure model}
\label{model}
Data \dat and model parameters  \m are linked by a physical model embodied by the forward operator $g(\cdot)$.

\begin{equation}
\dat = g(\m)
\label{equ_forward}
\end{equation}
For a given model $\m$ of the interior structure, mass $M$, radius $R$, and $\fesi$ are computed and compared with observed data $\dat$.
The function $g({\bf m})$ combines thermodynamic, Equation-of-State (EoS), and atmospheric modeling as described in the following sections.

\subsubsection{{\bf Iron core}}
In our model, we assume that the core is made of pure solid hcp (hexagonal close-packed) iron. Unlike in Earth's core, we neglect light elements and nickel and disregard
other iron polymorphs that stabilize at high temperatures. To compute the core density profile, we use an EoS for hcp  iron  provided by \citet{bouchet}.
It is based on results obtained from {\it ab initio} molecular dynamics simulations for pressures up to 1500 GPa and temperatures up to about 15000 K
and is in good agreement with experimental data obtained at Earth’s core conditions. This extensive pressure-temperature ($p$-$T$) range allows for modeling rocky exoplanets up to ten Earth masses (\ME).
Throughout, we assume an adiabatic temperature profile.

\subsubsection{{\bf Silicate mantle}}
Computing the mantle density profile is done in a manner analogous to \citet{dorn}.  Equilibrium mineralogy and density are computed as a function of pressure, temperature, and bulk composition by Gibbs energy minimization \citep{connolly09} within the model chemical system Na$_2$O-CaO-FeO-MgO-Al$_2$O$_3$-SiO$_2$. For these calculations the pressure is obtained by integrating the load from the surface boundary condition. As in \citet{dorn} we fix the thermal gradient in the mantle based on the adiabatic gradient of Earth's mantle. The mantle surface temperature equals the maximum of either the temperature at the bottom of the water layer or 1600~K (usual reference temperature of the Earth). For this purpose, we adopt the thermodynamic formulation of \citet{stixrudea} and parameters given in \citet{stixrudeb}.

\subsubsection{{\bf Water layer}}
Water has a rich phase diagram with a variety of structural transitions depending on temperature and pressure \citep[e.g.,][]{French}. In most of our planet realizations, temperatures in the water layer generally range from $\sim$250~K to $\sim$1000~K and pressures up to a few hundred GPa. 
In order to compute the density profile of the water layer, we follow \citet{Vazan}, using a quotidian equation of state (QEoS), which combines the Cowan ion EoS with the Thomas-Fermi model for electrons and treats H$_2$O as a mixture of atoms. This QEoS is in good agreement with the widely used ANEOS \citep{Thompson} and SESAME EoS \citep{Lyon}. 
Above 44.3 GPa, we use the tabulated EoS from \citet{seager2007} that is derived from DFT simulations and predict a gradual transformation from ice VIII to X. We assume an adiabatic thermal profile in the ice layer. 

\subsubsection{{\bf Atmospheric models}}
\label{atmomodel}
{\coco Previous works on mass-radius relationships are often restricted to pure H/He envelopes \citep[e.g.,][]{rogers2010, howe}. However, the compositional diversity might be large \citep{newsom} and significantly effect radius \citep[e.g.,][]{baraffe, vazan15}. Here, we employ two different atmospheric models that account for enriched atmospheres (with the caveat of assuming ideal gas behavior). }
Model I solves the radiative transfer equation. This model assumes ideal gas behavior and accounts for the presence of H, He, C, and O. It considers opacities that are adapted to solar abundances \citep{lodders03}.
More detailed and complex calculation of absorption and emission coefficients that inherit self-consistent opacities, scattering, clouds, and non-equilibrium chemistry could theoretically also be taken into account. However, in practice, the sparseness of available data does not warrant a more sophisticated treatment. Mass and radius observations will only allow us to constrain key characteristics of the envelope. 
For comparison, we also employ a second atmospheric model II that calculates an isothermal atmosphere with a simple pressure model using the scale-height model.  {\coco Model II is computationally very inexpensive. The validity of models I and II is  roughly restricted to 0.01 > \menv/$M$ and 0.0001 > \menv/$M$, respectively. Details on these limits are discussed in Section \ref{atmosection}. 
Both models are described in the following.}

\paragraph{{\bf Atmospheric model I}}

relies on the atmospheric code presented in \citet{VENTURINI2015}, which has been adapted to compute planetary radii. For a radius and mass of the solid interior, distance to star $a$, stellar effective temperature $T_{\rm star}$, stellar radius $R_{\rm star}$, {\coco planet envelope} luminosity $L$, envelope metallicity \Zenv, and envelope mass \menv, we solve the equations of hydrostatic equilibrium, mass conservation, and energy transport. As in \citet{VENTURINI2015}, we implement the CEA (Chemical Equilibrium with Applications) package \citep{CEA} for the EoS, which performs chemical equilibrium calculations for an arbitrary gaseous mixture,  including dissociation and ionization and assuming ideal gas behavior. We assume an envelope with an elemental composition of H, He, C, and O. We define the envelope metallicity as the mass fraction of C and O in the envelope, which can vary between 0 and 1. The reason to implement CEA and not a more sophisticated EoS (for example, one that can take into account degeneracy of free electrons) is simply because no such EoS exists for an arbitrary mixture of H, He, C, and O. 

These chemical elements are fundamental because they allow for the formation of key atmospheric molecules such as H$_2$O, CH$_4$, CO$_2$, and CO \citep{Madhusudhan2012, LODDERS2002, VISSCHER2011,heng}. Moreover, {\coco effects of electron degeneracy pressure} are important to compute radius of planets with massive envelopes. Even for the most extreme model realizations in this study where the mass fraction of the envelope is about 1~\% (for a planet of 7 \ME), we expect the error to be less than 10~\% in radius.

{\coco
For the energy transport, we adopt the model presented in  \citet{JIN2014}, where an irradiated atmosphere is assumed at the top of the gaseous envelope and for which the analytic irradiation model of \citet{GUILLOT2010} is adopted. This irradiation model assumes a semi-gray, globally averaged temperature profile. 
 Specifically we are using an analytical solution of the radiative transfer equation in the two-stream approximation.
This irradiation model assumes a semi-gray, global temperature-averaged profile \citep{GUILLOT2010}, for which optical depth $\tau$ is related to the infrared mean opacity  ($\kappa_{\text{th}}$) by
$
\nicefrac{d \tau}{dr} = \kappa_{\text{th}} \rho \,,
$
where $\rho$ is density.

For the temperature gradient of the irradiated atmosphere, we solve the radial derivative of Eq. 49 of \citet{GUILLOT2010}:

\begin{equation}
\begin{aligned}
T^4 = & \frac{3 T_{\text{int}}^4 }{4} \bigg[\frac{2}{3} + \tau \bigg] + \frac{3 T_{\text{eq}}^4 }{4} \bigg[\frac{2}{3} + \frac{2}{3 \gamma}\bigg\{1 + \bigg(\frac{\gamma \tau}{2} -1\bigg) e^{-\gamma \tau} \bigg\} \\ 
& + \frac{2\gamma}{3} \bigg( 1 - \frac{\tau^2}{2} \bigg) E_2(\gamma \tau) \bigg] \, ,
\end{aligned}
\end {equation}
where $\gamma = \kappa_{\text{v}}/ \kappa_{\text{th}}$ is the ratio between visible and infrared opacity, $T_{\text{int}}$ is the intrinsic temperature given by $T_{\text{int}} = (L/(4\pi \sigma R_P^2))^{1/4}$, and $E_{2} (\gamma \tau)$ is the exponential integral, defined by $E_{n}(z) \equiv \int_1^{\infty} {t^{-n} e^{-zt} dt}$ with $n = 2$.
The boundary between the irradiated atmosphere and the envelope is set at $\gamma \tau = 100 / \sqrt(3)$ \citep{JIN2014}. For $\gamma \tau$ larger than this, the usual Schwarzschild criterion to distinguish between convective and radiative layers is applied. That is, if the adiabatic temperature gradient is larger than the radiative one, the layer is stable against convection, and the radiative diffusion approximation is used for computing the temperature gradient:

\begin{equation}
\frac{dT}{dr} = - \frac{3 \kappa_{\text{th}} L \rho}{64 \pi \bar{\sigma}  T^3 r^2} \,\, ,
\end{equation}
where $L$ is the intrinsic luminosity, $\bar{\sigma}$ is the Stefan-Boltzmann
constant. Since we do not perform evolutionary calculations, $L$ is a model parameter (see Section \ref{parametrization}). However, when the radiative gradient is larger than the adiabatic gradient, the layer is convective, and the temperature gradient is assumed to be adiabatic (which is computed with the EoS).

In \citet{GUILLOT2010},  $\kappa_{\text{th}}$ and  $\kappa_{\text{v}}$  (and therefore, $\gamma$) are free parameters. In order to reduce the number of free parameters, we use the prescription of \citet{JIN2014} who calibrate $\gamma$ for different equilibrium temperatures in order to reproduce results from more sophisticated atmospheric models for which a wavelength-dependent opacity function is used while solving for radiative equilibrium \citep{PARMENTIER2013, FORTNEY2008}. We implement this calibration in our numerical scheme, that is  we interpolate the values of $\gamma$ for a given equilibrium temperature from Table 2 of \citet{JIN2014}. In this way, without using detailed opacity calculations in the treatment of irradiation, we mimic the fundamental physics underlying atmospheric absorption and re-irradiation in a more simple (and numerically inexpensive) fashion.  In order to compare the transit radius of a model realization with the measured radius from primary transits, we follow \citet{GUILLOT2010} and evaluate where the chord optical depth $\tau_{ch}$ becomes 2/3.
}

\paragraph{{\bf Atmospheric model II}}

assumes a simplified atmospheric model with a thin, isothermal atmosphere in hydrostatic equilibrium and ideal gas behavior, which is calculated using the scale-height model. {\coco For a given pressure \Pbatm, mean molecular weight $\mu$, mean temperature (parameterized by $\alpha$), number of scale heights of opaque layers $N$ and a given solid interior we compute planet radius. 

The scale-height $H$ is the increase in altitude for which the pressure drops by a factor of $e$ and can be expressed by

\begin{equation}
H = \frac{T_{\rm atm} R^{*}}{g_{\rm batm} \mu },
\end{equation}

where $g_{\rm batm}$ and $T_{\rm atm}$ are gravity at the bottom of the atmosphere and atmospheric temperature, respectively. $R^{*}$ is the universal gas constant (8.3144598 J mol$^{-1}$ K$^{-1}$) and $\mu$ the mean molecular weight.
The pressure $p$ at a given depth $z$ is the result of weight of the overlying gas layers. The hydrostatic equilibrium equation gives:

\begin{equation}
\frac{dp}{dz} = -gp .
\end{equation}
With the assumption that gravity $g$ is constant and using the EoS for ideal gas, the density $\rho$ can be expressed as:

\begin{equation}
\rho = \frac{p R^{*}}{T_{\rm atm} \mu}.
\end{equation}
The combination of the previous equations and the subsequent integration over pressure and altitude $z$ ($z=0$ where $p = p_0$ and $\rho = \rho_0$) leads to
$
p = p_0 \exp(-\nicefrac{z}{H})
$
and $
\rho = \rho_0 \exp(-\nicefrac{z}{H}).
$

The mass of the atmosphere $m_{\rm atm}$ is directly related to the pressure $p_{\rm batm}$ as:

\begin{equation}\label{massgas}
m_{\rm atm}= 4\pi  p_{\rm batm} \frac{r_{\rm batm}^2}{g_{\rm batm}}.
\end{equation}
where $r_{\rm batm}$ and $p_{\rm batm}$  are radius and pressure at the bottom of the atmosphere, respectively.  The thickness of the opaque atmosphere layer $z_{\rm atm}$ is:

\begin{equation}
z_{\rm atm}= H N,
\end{equation}
where $N$ is the number of opaque scale-heights $H$.
The atmosphere's constant temperature is defined as

\begin{equation}
T_{\rm atm}  = \alpha T_{\rm star} \sqrt{\frac{R_{\rm star}}{2 a}},
\end{equation}
where $R_{\rm star}$ and $T_{\rm star}$ are radius and effective temperature of the host star and $a$ is semi-major axes. {\coco The factor $\alpha$ is a model parameter (see Section \ref{parametrization}) and incorporates possible cooling and heating of the atmosphere, it can vary between 0 and $\alpha_{\rm max}$. There is an upper bound $\alpha_{\rm max}$, because there is a physical limit to the amount of warming by greenhouse gases. We approximate $\alpha_{\rm max}$ for a moist (water-saturated) atmosphere (see Appendix \ref{tlimit}).}

Generally, atmospheres can contain trace elements present at low pressures that have negligible contribution to the mass of the envelope but a significant contribution to the optical depth. In order to account for such effects, we use \Pbatm and $N$ as independent parameters.

 We have chosen to make model II very general, that is we decouple structure and transmissivity of the gas layer by distinguishing between $\mu$ and $N$. The equivalent procedure of this in model I would be to define opacities as free parameters. Model II has four compared to three degree of freedom in model I.
}

\begin{table*}[ht]
\caption{Prior model parameter ranges. \label{tableprior}}
\begin{center}
\begin{tabular}{lllc}
\hline\noalign{\smallskip}
parameter & prior range & distribution & model \\
\noalign{\smallskip}
\hline\noalign{\smallskip}
$r_{\rm core}$         & 0.01$ r_{\rm solid}$ -- 1 $r_{\rm solid}$ &uniform in $r_{\rm core}^3$& I, II\\
$\fesima$           & 0 -- $\fesistar$&uniform& I, II\\
$\mgsima$         & $\mgsistar$ &Gaussian& I, II\\
$r_{\rm solid}$   & 0.01$R$ -- 1.1 $R$& uniform & I, II\\
$m_{\rm water-ice}$ & 0 -- 0.98 $M$& uniform & I, II\\
\menv            & {\coco $10^{-10}$ \ME} -- 0.9 $M$ & {\coco uniform in log(\menv) }& I\\
$L$                & $10^{18} - 10^{23}$ erg/s& {\coco uniform in log($L$)}& I \\
\Zenv             & 0 -- 1& uniform in 1/\Zenv & I\\
\Pbatm  & 10$^{-4}$ -- 10$^9$ Pa & {\coco uniform in log(\Pbatm)}& II\\
$N$                    & 0 -- log(10$^9$/10$^{-4}$) $\approx$ 30 &uniform & II\\
$\mu$                 & 2.3 -- 50.0& uniform in 1/$\mu$&II \\
$\alpha$             & 0.0 -- $\alpha_{\rm max}$ &uniform & II\\
\hline
\end{tabular} 
\end{center}
\end{table*}

\subsection{Prior information}
\label{prior}

Table \ref{tableprior} lists prior parameter distributions. {\coco The chosen prior parameters distributions are wide reflecting a conservative choice. Different priors are discussed in Section \ref{prima}.}

  Prior bounds on $\fesima$ and $\mgsima$ are linked to the stellar abundance constraints. Since all Si and Mg are assumed to be in the mantle, $\mgsistar$ defines the prior on $\mgsima$. We assume $\mgsistar$ to be Gaussian distributed. Fe, on the other hand, is distributed between core and mantle. Thus, the bulk abundance constraint $\fesi$ \mbox{(= $\fesistar$)} defines only the upper bound of the prior on $\fesima$. There is an additional numerical limitation that the absolute iron oxide abundance in the mantle cannot exceed 70~\%.
For \Pbatm (model II), \menv and $L$ (model I), we assume {\coco the logarithm of these parameters to be uniformly distributed}. The upper bound on the mass of the envelope in model I is set to 90~\% of the planet mass, which is roughly the scale of Saturn and possibly Jupiter.  {\coco The range of luminosities $L$ is chosen such that it embraces those of the Moon and Neptune.}
For model II, the mass of the envelope is parameterized through \Pbatm. Its prior upper bound is arbitrarily set to 1 GPa. At such high pressures, the atmosphere  may no longer behave like a gas and the simplified pressure scale-height model becomes invalid \citep[e.g.,][]{Andrews}. Only model realizations with \Pbatm well below 1 GPa can be used for further interpretation. The temperature-related parameter $\alpha$ uniformly varies between 0 and $\alpha_{\rm max}$, making up for possible cooling and heating of the atmosphere; $\alpha_{\rm max}$ scales with surface gravity (see Appendix \ref{tlimit}).

   {\coco   An example of the influence of different priors on interior model predictions is discussed at the end of this study. Some examples are also shown in \citet{rogers2010}. In a future study, we will address this problem in more detail.}

\section{Results}
\label{Results}

\begin{figure*}[ht]
\centering
 \includegraphics[width = .8\textwidth, trim = 1cm 0cm 0cm 0cm, clip]{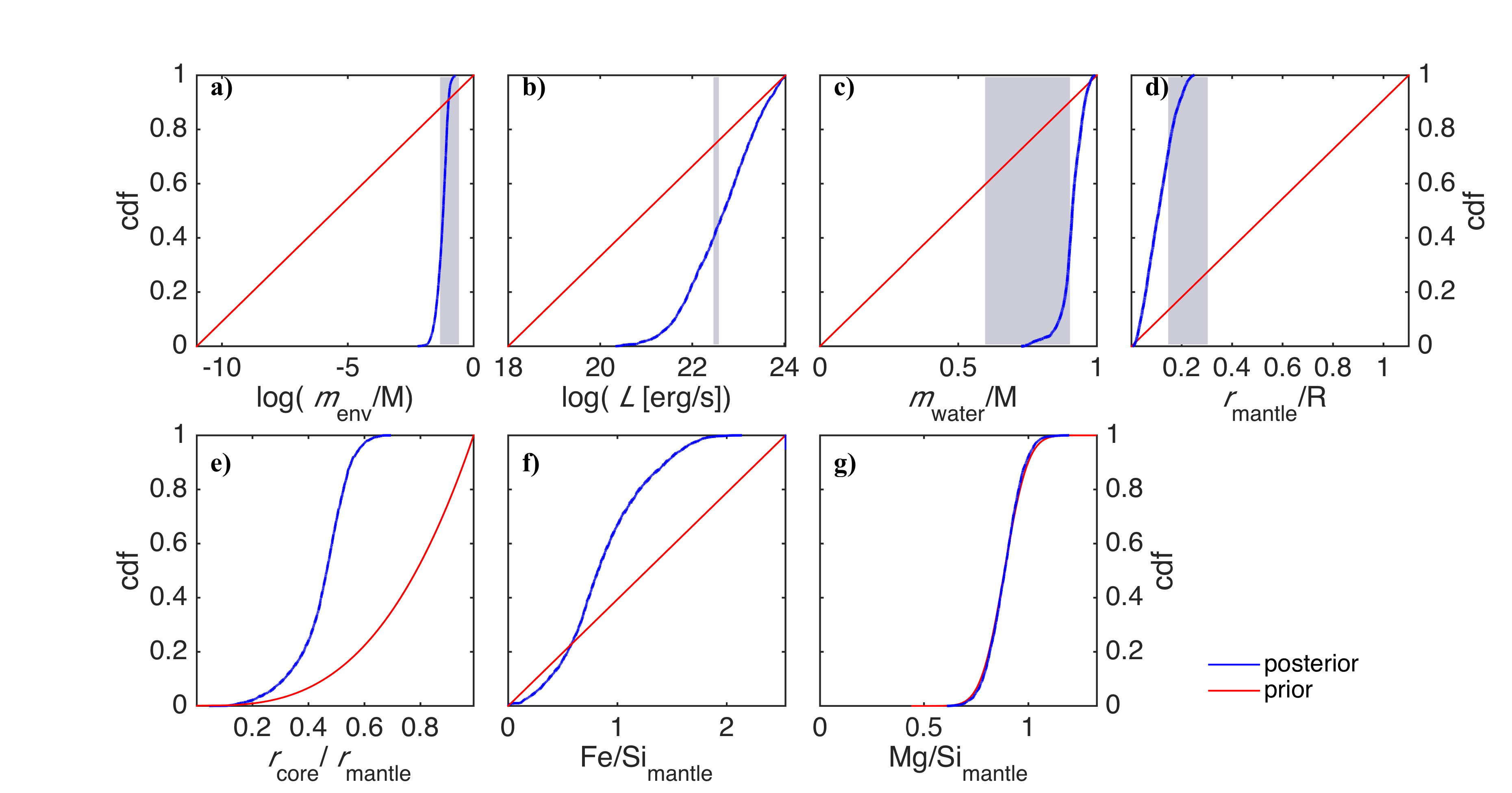}\\
 \caption{Sampled one-dimensional (1-D) marginal posterior cdfs (blue) of model I parameters for Neptune:
(a) mass of envelope \menv, (b) envelope Luminosity $L$, (c) mass of water \mice, (d) mantle radius \rsolid, (e) core radius \rc, (f) $\fesima$, (g) $\mgsima$. Prior and posterior nearly completely overlap in (g). The envelope metallicity \Zenv (not shown) is fixed, \Zenv = 0. The prior cdfs are plotted in red. {\coco Gray area in plots (a-d) represent independent literature estimates (see main text).} \label{NepGas}}
\end{figure*}

\begin{figure}[ht]
\centering
 \includegraphics[width = .5\textwidth, trim = 1cm 0cm 1cm 0cm, clip]{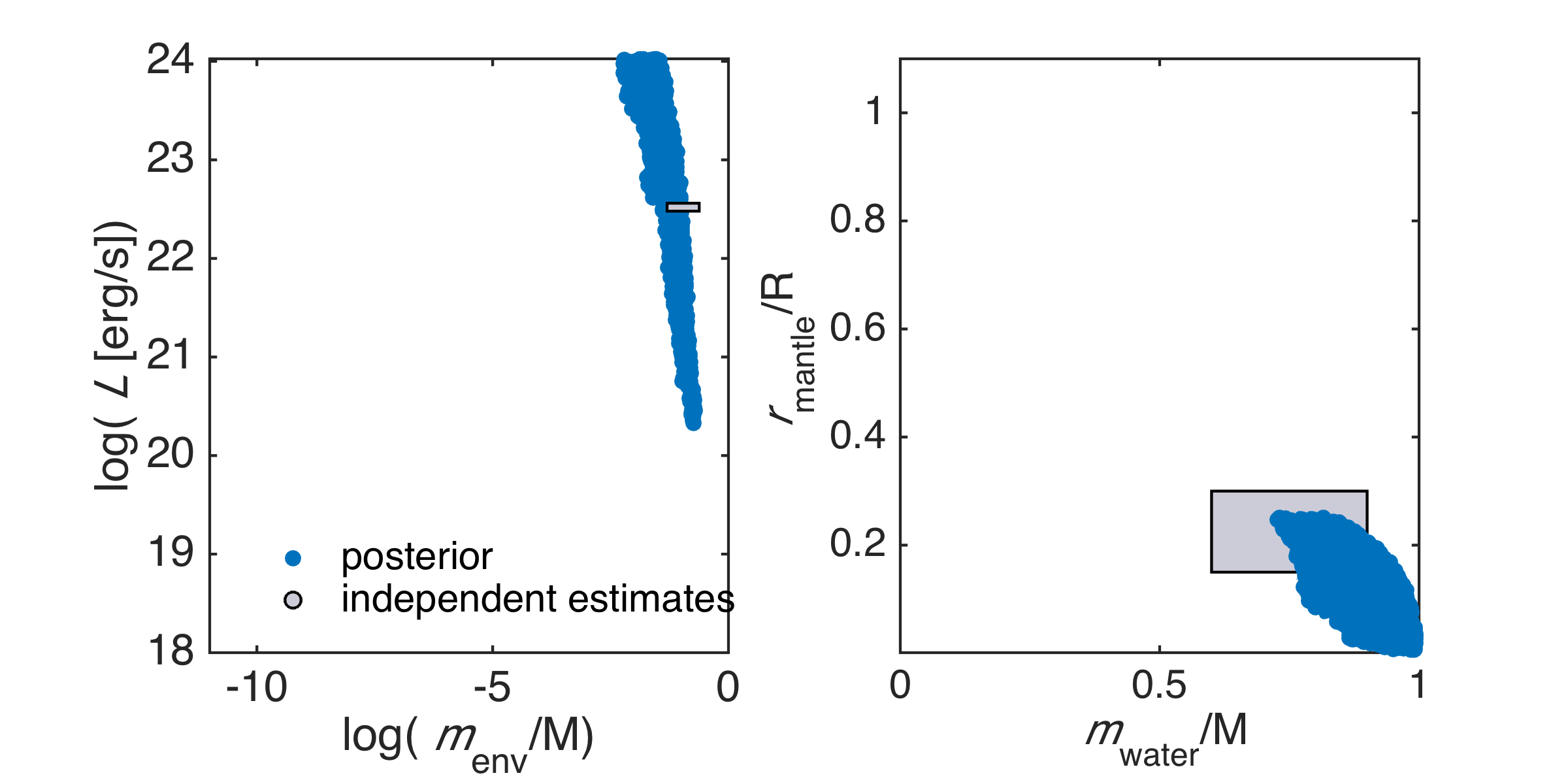}\\
 \caption{{\coco Sampled two-dimensional (2-D) marginal posterior pdfs (blue) of model I parameters for Neptune:
(a) mass of envelope \menv and envelope Luminosity $L$, (b) mass of water \mice and mantle radius \rsolid.  {\coco Gray areas represent independent literature estimates (see main text)}.} \label{Nep2D}}
\end{figure}

\begin{figure*}[ht]
\centering
 \includegraphics[width = .95\textwidth]{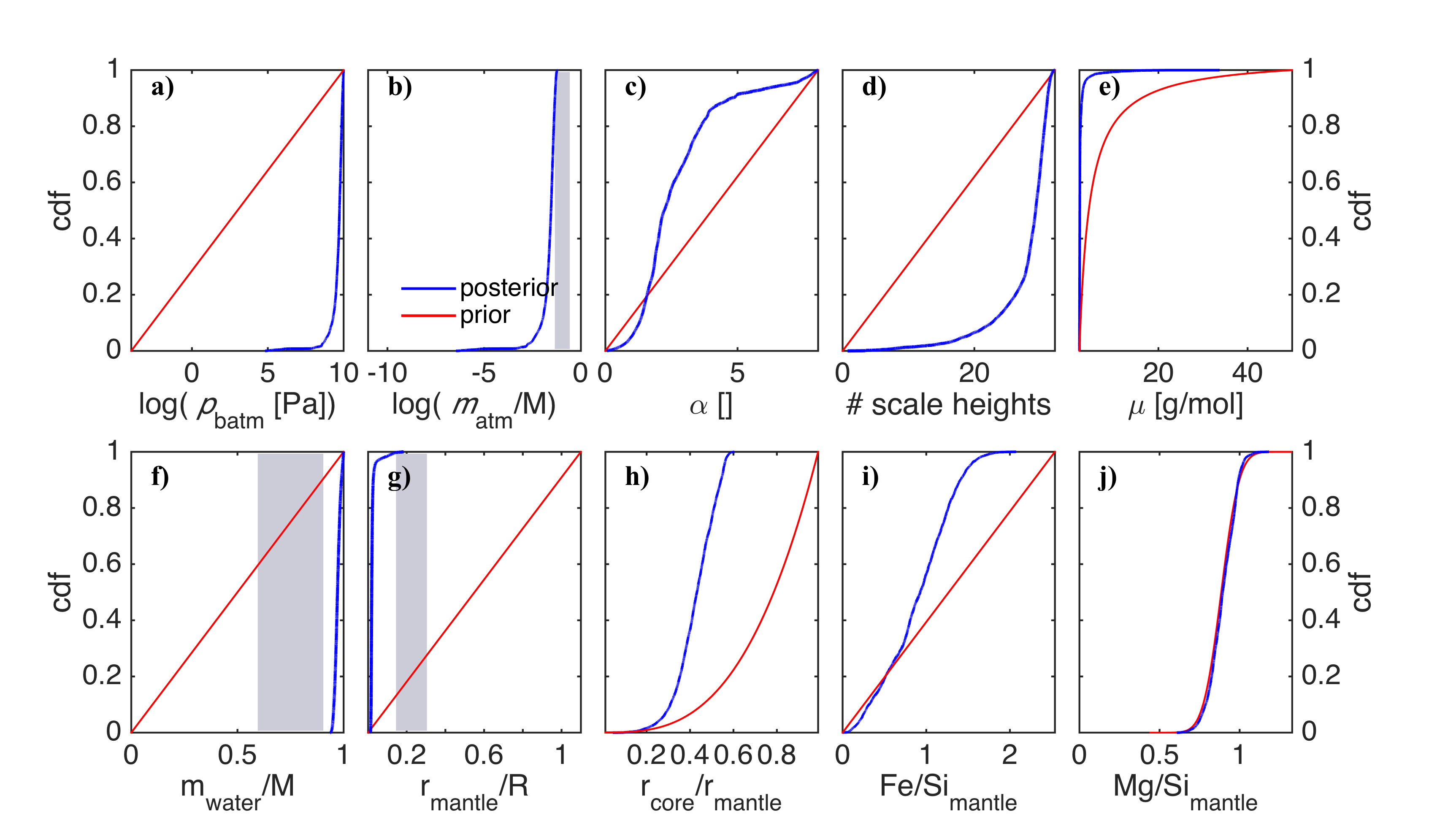}\\
 \caption{Sampled 1-D marginal posterior cdfs (blue) of model II parameters for Neptune:
(a) pressure at bottom of atmosphere \Pbatm, (b) atmospheric mass fraction m$_{\rm atm}$/M (Eq. \ref{massgas}), (c) temperature-related parameter $\alpha$, (d) number of scale-heights of opaque layers $N$, (e) mean molecular weight $\mu$, (f) mass of water \mice, (g) mantle radius \rsolid, (h) core radius \rc, (i) $\fesima$, (j) $\mgsima$. The prior cdfs are plotted in red.  {\coco Gray areas in (b,e,f) represent independent literature estimates (see main text).} \label{NepScale}}
\end{figure*}

\subsection{Method validation: Neptune}
\label{validation}

As in \citet{dorn}, we validate the methodology against solar system planets. Here, we compare with Neptune ($M = $17.15 \ME, \mbox{$R = 3.87$ \RE},  {\coco where \RE is 1 Earth radius}), the smallest volatile-rich solar system planet. For model I, we have restricted the gas envelope to a pure H/He gas layer (\Zenv = 0) and use the more appropriate EoS of \citet{saumon} for Neptune, since the (otherwise employed) assumption of ideal gas behavior can result in radius uncertainties larger than 10~\% for a gas mass fractions of a few percent. Although both atmospheric models I and II are not specifically tailored for Neptune, their application serve as a benchmark test and are not meant to provide new insights on Neptune's interior. 


For Neptune, geophysical data (gravitational and magnetic moments, solid-body rotation period, and heat flux) and atmospheric composition estimates are available that provide us with constraints on a possible three-component interior: (1) an outermost molecular envelope largely composed of H/He, (2) a weakly conducting ionic ocean of water, methane, and ammonia, and (3) a rocky central core \citep[e.g.,][]{soderlund, podolak, ness}. The transition between outermost envelope and ocean is predicted to be around 0.8~$R$ by \citet{lee}, whereas the transition from ocean to rock likely occurs below 0.3~$R$ \citep{redmer}. The transitions are neither well determined \citep{podolak, Nettelmann} nor necessarly sharp \citep{helled}.
For a three-component structure of H/He, H$_2$O, and SiO$_2$, \citet{helled} suggest an upper bound on the water mass fraction of 90~\% and an upper bound on the envelope mass fraction of 24~\%. If the ice/rock ratio is restricted to proto-solar, \citet{hubbard} find that Neptune could consist of about 25~\% rock, 60-70~\% ice, and \mbox{5-15~\%} gas by mass.

Here, we use  uncertainties of $1~\%$ on both the observed $M$ and $R$, and $10~\%$ on the solar ratios $\fesistar$ and $\mgsistar$ \citep{lodders03}. Results for the two atmospheric models are shown in Figures \ref{NepGas} and \ref{NepScale}, respectively. The one-dimensional (1-D) marginal posterior cumulative distribution function (cdf) for each model parameter (in blue) is plotted with the prior distribution (in red) and independent parameter estimates (gray areas). The cdf describes the probability of a model parameter \m with a certain probability distribution to be less or equal to a given value of \m. In addition, Figure \ref{Nep2D} shows the 2-D marginal posterior pdfs for those model parameters of model I for which we have independent estimates.
These plots suggest the following:
\begin{itemize}
\item {\coco The interior structure of Neptune is constrained by the data.}
\item Available independent parameter estimates (shown in gray) overlap with the blue posterior cdfs for \menv, $L$, \mice, and \rsolid (model I, Figures \ref{NepGas} and \ref{Nep2D}); for model II (Figure \ref{NepScale}) this is only the case for $m_{\rm atm}$ (derived from  \Pbatm and Eq. \ref{massgas}), \mice and \rsolid are over-and under-predicted, respectively.
\item With only mass, radius, and abundance constraints, our method (model I) predicts independent geophysical estimates of Neptune's interior. Compared to independent estimates, our calculated confidence regions for the structural parameters are larger, since we rely on limited data:
 \begin{itemize} \itemsep1pt
   \item[] 0.01 $<$ \menv/$M$ $<$ 0.2,
  \item[] 0.75 $<$ \mice/$M$ $<$ 0.98,
 \item[] 0.01$<$ \rsolid $<$ 0.25,
  \item[] $10^{21}$ erg/s$< L< 10^{24}$ erg/s.
 \end{itemize}
\item The simplified pressure model II leads to an overestimation of \mice and {\coco underestimation} of \rsolid compared to model I. This is because the same radius fraction of gas results in different $p$-$T$ boundary conditions for the ice layer for both models. The simplified pressure model II generally overestimates \Pbatm, which leads to an increase in water ice density. In order to fit the radius, the higher water ice density implies a larger \mice. At the same time, the mass contribution of the rocks needs to be reduced so as not to overestimate mass. 

\end{itemize}

{\coco Without the restriction to pure H/He in model I and under the assumption of ideal gas, the results are similar with the largest discrepancy in the estimate of a gas mass fraction (with a 50\%-percentile of 0.01 \menv/M under the ideal gas premise compared to 0.06 \menv/$M$ in Figure \ref{NepGas})}.


\begin{table*}[Ht]
 \caption{Data of synthetic planets. \label{tabledata}}
\begin{tabularx}{1.03\textwidth}{@{} l |cccccc >{\setlength{\baselineskip}{0.8\baselineskip}}Y c l@{}} 
\hline\noalign{\smallskip}
name & $M$ [\ME] & $\sigma_M$ & $R$ [\RE] & $\sigma_R$& $\sigma_{\fesi}$ &$\sigma_{\mgsi}$  & semi-major \newline axis [AU]&$\bar{\rho}$ [g/cm$^3$]& additional comments\\
\hline\noalign{\smallskip}
Case A& 7 & 5~\% & 1.7 & 2~\% & 20~\% & 20~\% & 1 &7.86 & Figs. \ref{VaryRadii1}, \ref{VaryRadii2}\\
Case B& 7 & 5~\% & 2.2 & 2~\% & 20~\% & 20~\% & 1 & 3.62&Figs. \ref{VaryRadii1}, \ref{VaryRadii2}, \ref{VaryScale}, \ref{VaryGas}\\
Case C& 7 & 5~\% & 2.6 & 2~\% & 20~\% & 20~\% & 1 &2.20& Figs. \ref{VaryRadii1}, \ref{VaryRadii2}, \ref{HHe1}, \ref{Comp1}, \ref{Comp2}\\
Case D& 7 & 5~\% & 2.9 & 2~\% & 20~\% & 20~\% & 1 &1.58 &Figs. \ref{VaryRadii1}, \ref{VaryRadii2}\\
Case E& 7 & 20~\% & 2.2 & 10~\% & 20~\% & 20~\% & 1 &3.62& Figs. \ref{VaryScale}, \ref{VaryGas}\\
Case F& 7 & 5~\% & 2.2 & 2~\% & 50~\% & 50~\% & 1 &3.62& Figs. \ref{VaryScale}, \ref{VaryGas}\\
Case G& 7 & 5~\% & 2.2 & 2~\% & 80~\% & 80~\% & 1 &3.62& Figs. \ref{VaryScale}, \ref{VaryGas}\\
Case H& 7 & 5~\% & 2.6 & 2~\% & 20~\% & 20~\% & 0.1 &2.20& Fig. \ref{HHe1} \\
Case J& 7 & 5~\% & 2.6 & 2~\% & 20~\% & 20~\% & 0.1 &2.20& H/He atmosphere only, Fig. \ref{HHe1}\\
Case K & 7 & 5~\% & 2.6 & 2~\% & 20~\% & 20~\% & 1 &2.20& H/He atmosphere only, Fig. \ref{HHe1}\\
\end{tabularx}%
\end{table*}

\begin{figure*}[Ht]
\centering
 \includegraphics[width = .7\textwidth]{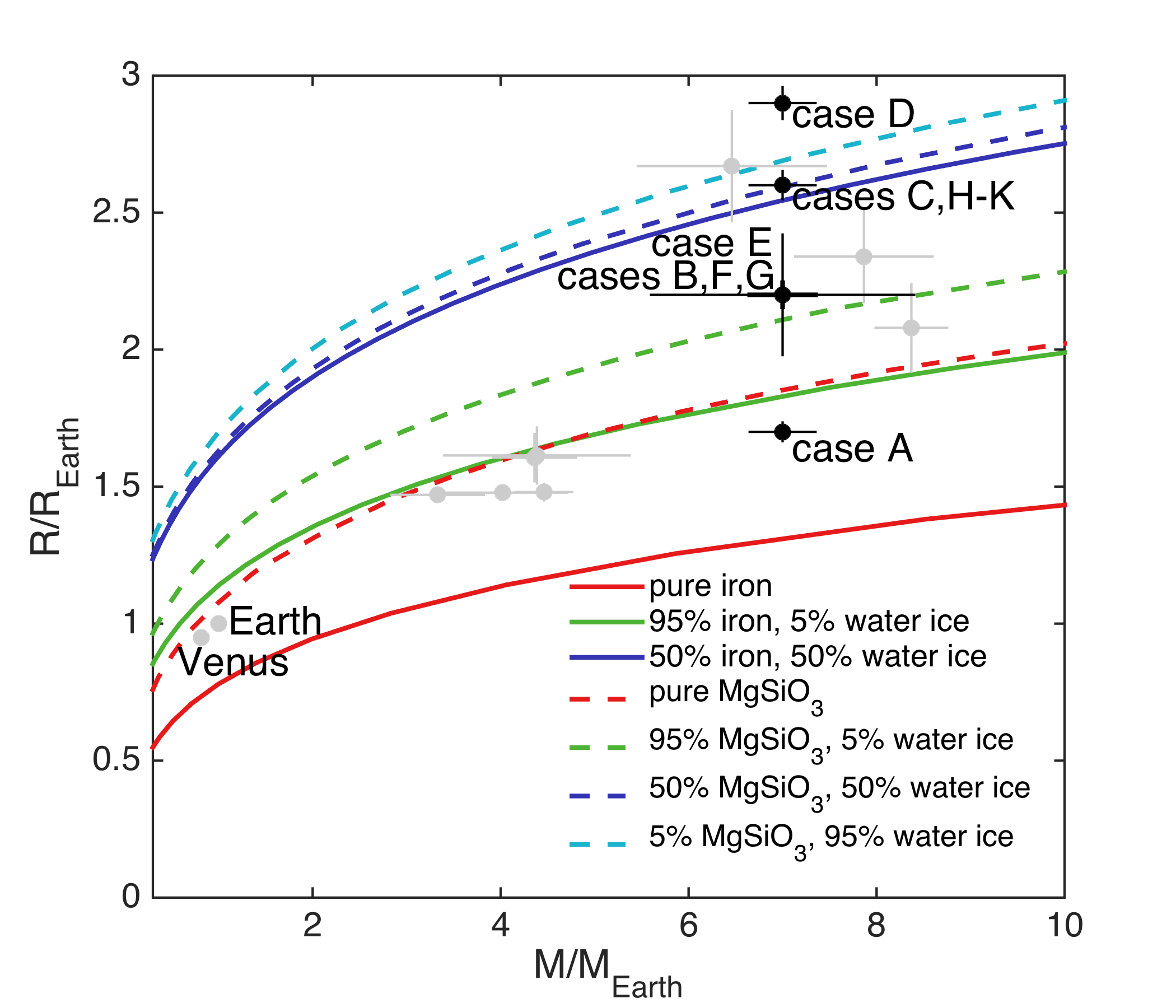}\\
 \caption{Masses and radii of synthetic planets (black dots, cases A-K), observed exoplanets (gray dots) from \citet{dressing}, and Earth and Venus. Planets are plotted against mass-radius curves of idealized compositions for which a surface temperature of 300 K has been assumed. Planet cases A-K are summarized in Table \ref{tabledata}. \label{MR_1}}
\end{figure*}

\subsection{Synthetic cases}

Next, we apply our method to synthetic exoplanets. Application to actual observations is presented in a companion paper \citep{dornA}. In this study, we emphasize instead the influence of the following parameters on interior predicitions: bulk density $\bar{\rho}$, data uncertainties, semi-major axis, atmospheric composition, {\coco and prior distributions}. For the latter, we test the a priori assumption of enriched envelopes versus pure H/He envelopes. For all synthetic planets we assume $M = 7$ \ME, since the transition between rocky and non-rocky planets seems to occurr around this mass \citep[e.g.,][]{weiss11, rogers15}. Table \ref{tabledata} lists all relevant data for the synthetic cases and Figure \ref{MR_1} shows their masses and radii plotted against curves of idealized compositions. For all synthetic cases, we assume solar values for abundance constraints \citep{lodders03}, stellar effective temperature and stellar radius of the Sun. 
In the following, we discuss these test cases.

\begin{figure*}[ht]
\centering
 \includegraphics[width = .8\textwidth]{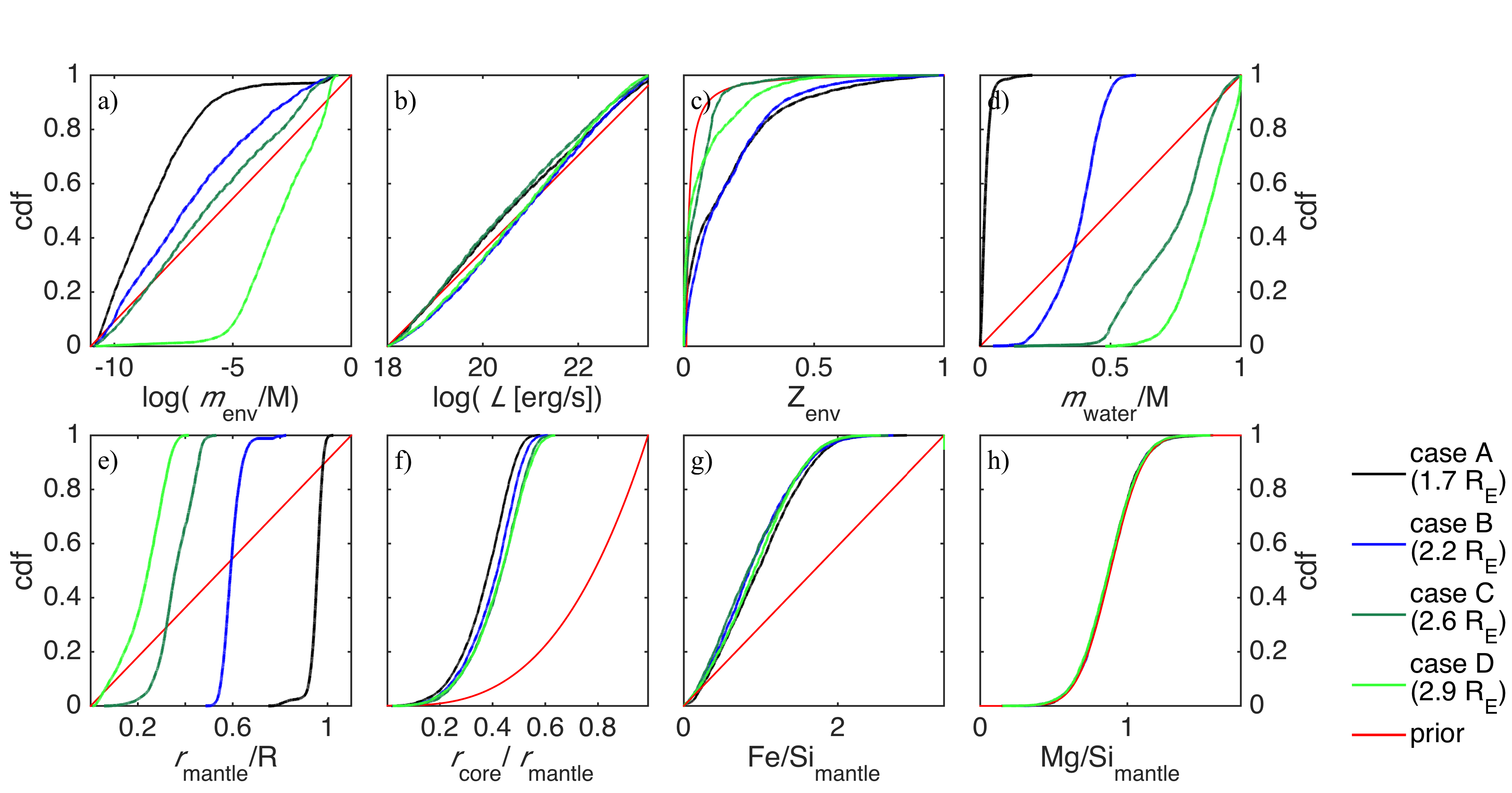}\\
 \caption{Sampled 1-D marginal posterior cdfs of model I parameters for synthetic planet cases (A-D) of 7 \ME that vary in terms of radii: \mbox{1.7 \RE (A)}, 2.2 \RE (B), 2.6 \RE (C), 2.9 \RE (D);
(a) mass of envelope \menv, (b) envelope luminosity $L$, (c) envelope metallicity \Zenv,
   (d) mass of water \mice, (e) mantle radius \rsolid, (f) core radius \rc, (g) $\fesima$, (h) $\mgsima$. \label{VaryRadii1}}
\end{figure*}

\begin{figure*}[Ht]
\centering
 \includegraphics[width = .85\textwidth]{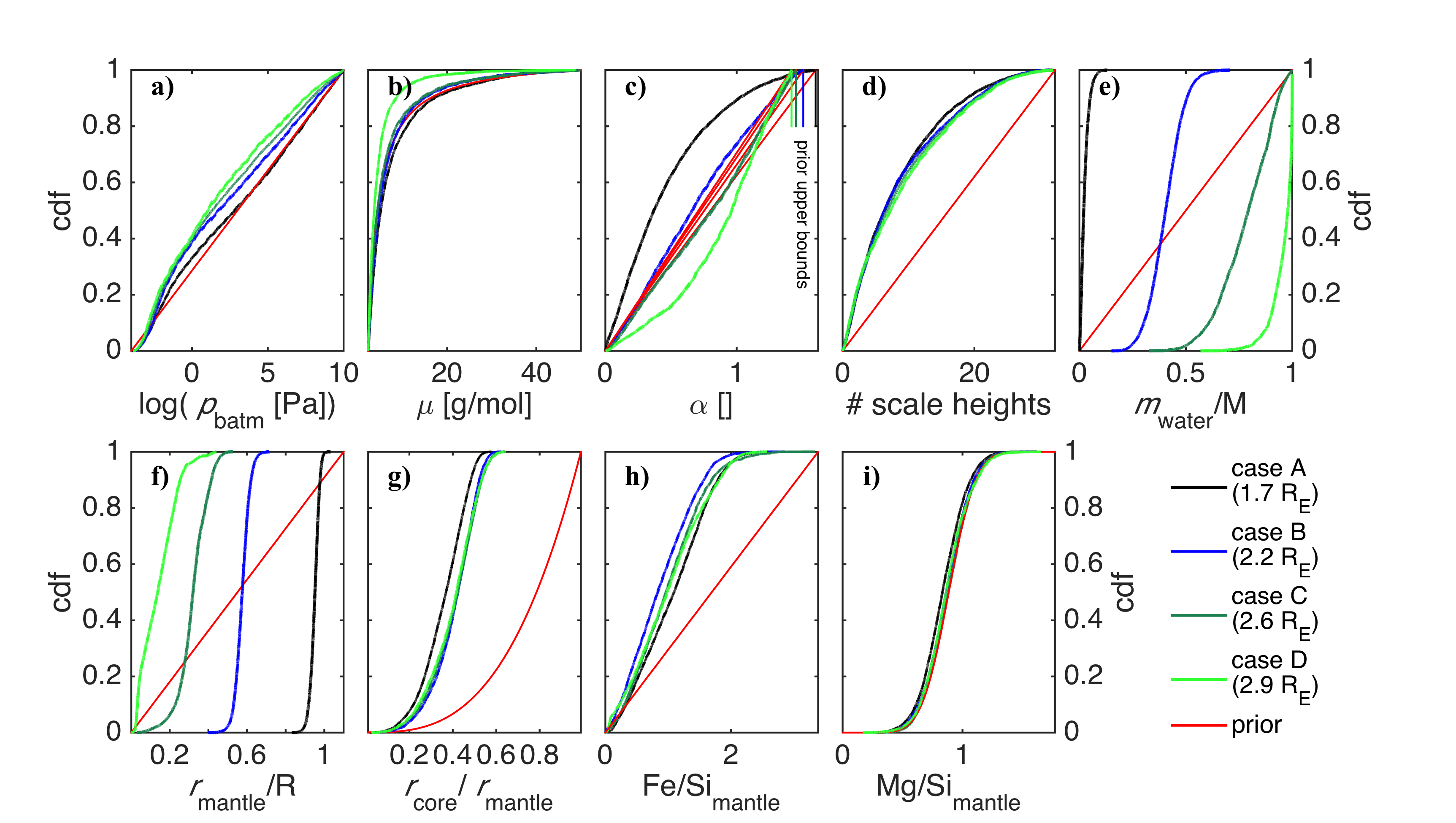}\\
 \caption{Sampled 1-D marginal posterior cdfs of model II parameters for synthetic planet cases  (A-D) of 7 \ME that vary in terms of radii: \mbox{1.7 \RE (A)}, 2.2 \RE (B), 2.6 \RE (C), 2.9 \RE (D);
(a) pressure at bottom of atmosphere \Pbatm, (b) atmospheric mean molecular weight $\mu$, (c) temperature-related parameter $\alpha$, (d) number of scale-heights of opaque layers $N$, 
   (e) mass of water \mice, (f) mantle radius \rsolid, (g) core radius \rc, (h) $\fesima$, (i) $\mgsima$. Depending on the case, the upper prior bound in (c) differs, which is indicated by the vertical colored lines corresponding to the respective case. \label{VaryRadii2}}
\end{figure*}

 \subsubsection{Influence of bulk density}
Planets A, B, C, and D are assigned different radii (1.7, 2.2, 2.6, and 2.9 \RE) and hence bulk densities $\bar{\rho}$ (Table \ref{tabledata}). Uncertainties for mass and radius are assumed to be similar to the predicted uncertainties from the PLATO mission \citep{plato}, that is 5\% and 2\%, {\coco respectively}.
 The influence of planet bulk density on retrieved parameters is shown in Figures \ref{VaryRadii1} and \ref{VaryRadii2}. 
 We observe, as expected, that bulk density {\coco correlates positively} with the size of the rocky interior \rsolid, and {\coco correlates negatively} with mass of water (\mice) and gas (\menv).
 Core size and mantle composition (Figure \ref{VaryRadii1}f-h and  \ref{VaryRadii2}g-i) show only small variations, because they are {\coco constrained by the solar abundances}. 

Among the parameters characterizing the gas layer for model I (Figure \ref{VaryRadii1}), \menv and \Zenv are constrained by data, whereas envelope luminosity $L$ is not. For the planet with the highest bulk density (case A) the gas layer contributes very little to planet radius, i.e., metallicity is high and/or \menv is small. Case A is found with a 90~\% probability to have an atmosphere smaller in mass than Earth ($10^{-7}$ \menv/$M$). {\coco Compared to high bulk density planets, low density planets can have gas of lower metallicity while gas mass fraction tends to be higher.} For very low density planets (case D) when even pure water ice is not sufficient to explain radius, small \menv are excluded as a result of which  \menv is larger than $10^{-5}$~$M$ with a probability of 90~\%.

The gas layer parameters for model II (Figure \ref{VaryRadii2}) indicate that the number of opaque scale-heights $N$ and temperature (parameterized by $\alpha$) in the gas layer appear to be best constrained by data. The expected trend of a higher temperature (larger $\alpha$) and an increased number of scale-heights that are needed to explain low bulk density planets is clearly visible (Figure \ref{VaryRadii2}c and d).
Mean molecular weight $\mu$ and \Pbatm are both weakly constrained for the high bulk density cases (A, B, and C).  When pure water ice cannot compensate enough to fit radius (case D compared to the other cases) the gas layer moves to higher pressures \Pbatm, lower mean molecular weights, higher temperatures ($\alpha$), and more scale-heights (Figure \ref{VaryRadii2} light green curve).

Although the use of both atmospheric models yield very similar parameter distributions for the rocky part of the planet, there are significant differences in \mice, particulary for the low density planets (cases C and D). This is because parameters related to gas and ice layers are those with the largest influence on planet radius. Hence differences in the atmospheric model affect the gas structure and in consequence the distribution of \mice. We will discuss these differences in more detail in the following.

\begin{figure*}[H]
\centering
 \includegraphics[width = 1.\textwidth, trim = 3.5cm 0cm 3.8cm 0cm, clip]{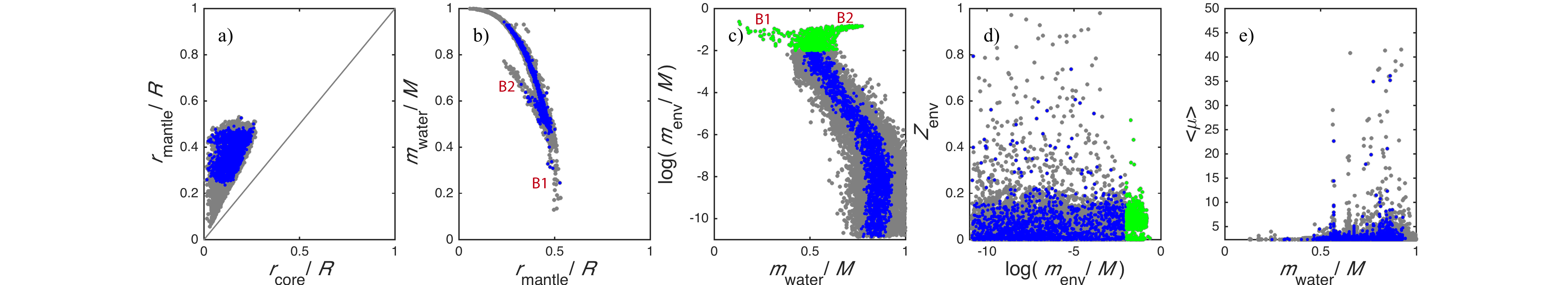}\\
 \caption{Sampled 2-D marginal posterior pdfs of model I parameters for synthetic planet case C showing the correlation between: (a) \rc and \rsolid, (b) \rsolid and \mice, (c) \mice and \menv, (d) \menv, and \Zenv, (e) \mice and the averaged $\mu$ corresponding to \Zenv. Those model realizations that explain the data within 1-$\sigma$ are plotted in blue. Samples in (c, d) for which gas mass fractions \menv/$M$ > 0.01 are highlighted in green and should be taken with care. See main text for discussion of features B1 and B2. \label{Comp1}}
\end{figure*}

\begin{figure*}[H]
\centering
 \includegraphics[width = 1.\textwidth, trim = 2cm 0cm 2cm 0cm, clip]{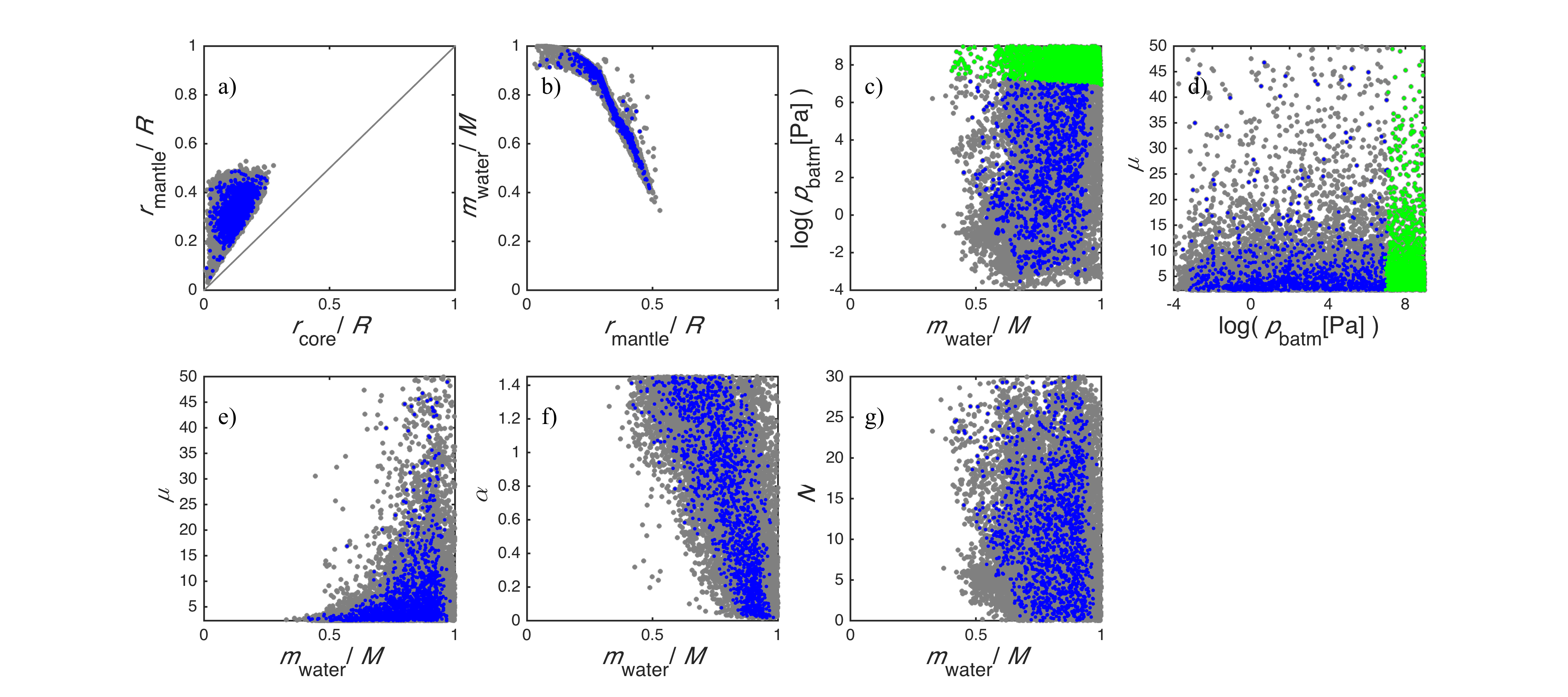}\\
 \caption{Sampled 2-D marginal posterior pdfs of model II parameters for synthetic planet case C showing the correlation between: (a) \rc and \rsolid, (b) \rsolid and \mice, (c) \mice and \Pbatm, (d) \Pbatm, and $\mu$, (e) \mice and $\mu$, (f)  \mice and $\alpha$, (g) \mice and $N$. Those model realizations that explain the data within 1-$\sigma$ are plotted in blue. Samples in (c, d) for which gas mass fractions \menv/$M$ > 0.0001 are highlighted in green and should be taken with care. \label{Comp2}}
\end{figure*}

 \subsubsection{Influence of atmospheric model}
 \label{atmosection}
Here, we take a closer look at the different parameter estimates for case C when using model I and II. We plot the sampled 2-D marginal posterior distributions of model parameters in Figures \ref{Comp1} and \ref{Comp2}. Overall, the distributions show similar trends with clear differences for the rocky and icy interior depending on atmospheric model:
\begin{itemize}
\item   There is a strong correlation between \mice and \menv in model I (Figure \ref{Comp1}). For model II, the  corresponding correlation between \mice and \Pbatm is weak. This reflects a higher degeneracy in the gas layer parameters for model II (more degrees of freedom).  
\item For model II, strongest correlations with \mice are seen for $\mu$ and $\alpha$ among the gas parameters.
\item For model I compared to model II, \rsolid tends to be larger (Figures \ref{Comp1}a and \ref{Comp2}a).
\item There is a clear discrepancy in the estimated \mice between the two models. For model I, the minimum \mice is estimated to be about 0.1 $M$, whereas for model II it is 0.5 $M$. 
\end{itemize}   
Model II leads to the misinterpretation that relatively low-density planets (case C) require a massive ocean to explain mass and radius. This is in line with earlier conclusions suggesting that it is impossible to distinguish between a thick atmosphere and an ocean based on mass and radius alone \citep[e.g.,][]{adams}. {\coco This is important in view of the different formation histories implied by either interpretation.} The results show that the simplified pressure model II fails to explain thicker atmospheres and thereby overestimates the amount of water ice. This is because it does not account for energy transport and thus overestimates the pressure increase with atmospheric depth.
Thicker atmospheres can in principle be realized, if temperatures (i.e., $\alpha$) exceeding the prior range ($\alpha_{max}$, Appendix \ref{tlimit}) would be allowed, implying a {\coco larger} greenhouse effect. 
However, there is a physical upper limit, the Komabayashi–Ingersoll Limit {\coco \citep{Komabayasi, ingersoll}}, to the amount of outgoing long-wave radiation that can be absorbed and emitted by greenhouse gases that warm the atmosphere.  More advanced modeling would be required to determine this upper limit for the studied cases, but this is outside of the scope of this study.

In the 2-D plots (Figure \ref{Comp1}b and c) showing the correlation between \rsolid and \rc, and \rsolid and \mice, respectively, two `branches' ({\coco labeled} B1 and B2) are visible (valid for massive atmospheres \menv $>$ 0.01 $M$ ) which are characterized by: 
\begin{itemize}\itemsep 2pt
\item B1:\\  \mice $<$ 0.5 $M$,\\ \Zenv$<$ 0.02, \\$L > 10^{22.5}$ erg/s
\item B2:\\ \mice $>$ 0.5 $M$, \\0.02 $<$\Zenv$<$ 1.0, \\ $10^{18}$erg/s $< L < 10^{22.5}$ erg/s
\end{itemize}
For gas envelopes of supersolar abundances (B2), self-gravity of massive gas layers leads to compressed envelopes. To fit radius in this case, a large \mice is required. For subsolar abundances and very high luminosities (B1), the envelopes are thick and make up for a large fraction of planet radius (> 25\%). However,  a minimum \mice of 0.1 $M$ appears to be required to fit radius. This is because we restrict the prior {\coco range on} luminosity $L$ to a maximum of $10^{23}$ erg/s (Neptune-like $10^{22.52}$ erg/s). If larger luminosities than the prior {\coco range} were allowed, thicker gas layers with negligible ice mass fractions could be realized. This suggests that constraints on the luminosities would allow to partly lift the degeneracy between an ocean and a thick atmosphere. This will be investigated in more detail in the future.

   {\coco We compare the planetary radii that are computed with both atmospheric models by using the calculated pressures and temperatures from model I (e.g., pressures at bottom and top of the gas layer and an averaged temperature) as input in model II for a rocky interior of 7 \ME. For an envelope mass of  \menv$ > 10^{-3}$ \ME (corresponding to \Pbatm $\approx$ 1000 bar), the discrepancy in radius becomes comparable to the observed radius uncertainty of 2~\%. We note that the comparison of both models is sensitive to the choice of temperature averaging.}
    Hence, for large bulk density planets with thin atmospheres (cases A and B), the choice of atmospheric model does not significantly affect  estimates of the rocky and icy interior (Figures \ref{VaryRadii1} and \ref{VaryRadii2}), whereas it becomes relevant for relatively low-density planets (cases C and D). 
   
     For the cases studied here, we conclude that the more accurate representation of gas layer physics makes model I more favorable inspite of larger computational costs.
   In the case of thin atmospheres, model II is valid. 


\begin{figure*}[h]
\centering
 \includegraphics[width = 1.\textwidth]{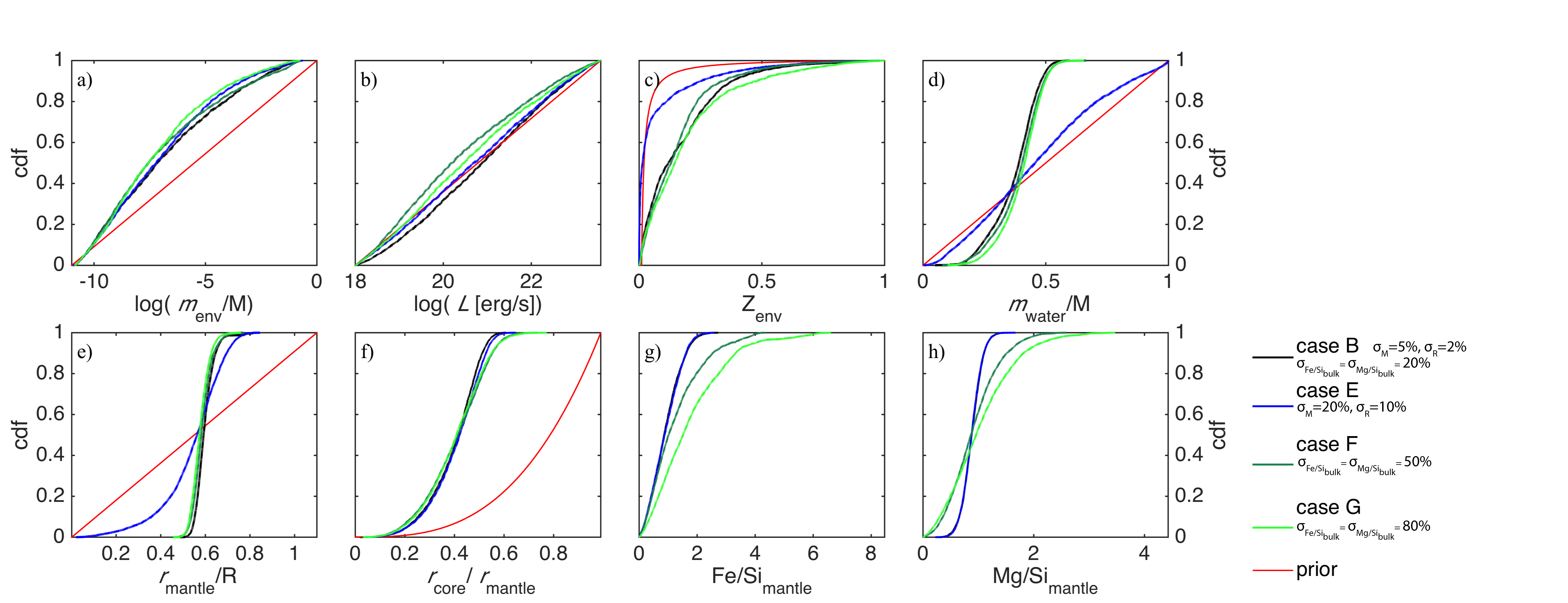}\\
 \caption{Sampled 1-D marginal posterior cdfs of model I parameters for synthetic planet cases B, E, F, G that vary in terms of data uncertainties. B is the reference case ($\sigma_M$=0.05~$M$, $\sigma_R$ = 0.02~$R$, 20~\% for both $\sigma_{\fesi}$ and $\sigma_{\mgsi}$), E has larger uncertainties in mass and radius ($\sigma_M$=0.2~$M$, $\sigma_R$ = 0.1~$R$), whereas F and G have larger uncertainties in the abundance constraints, 50~\% and 80~\%, respectively.
(a) Mass of envelope \menv, (b) envelope luminosity $L$, (c) envelope metallicity \Zenv,
   (d) mass of water \mice, (e) mantle radius \rsolid, (f) core radius \rc, (g) $\fesima$, (h) $\mgsima$.  The priors in (g) and (h) are not shown as not to overload the plot, because they differ among the cases. \label{VaryGas}}
\end{figure*}

\begin{figure*}[h]
\centering
 \includegraphics[width = .9\textwidth, trim = 0cm 0cm 0cm 0cm, clip]{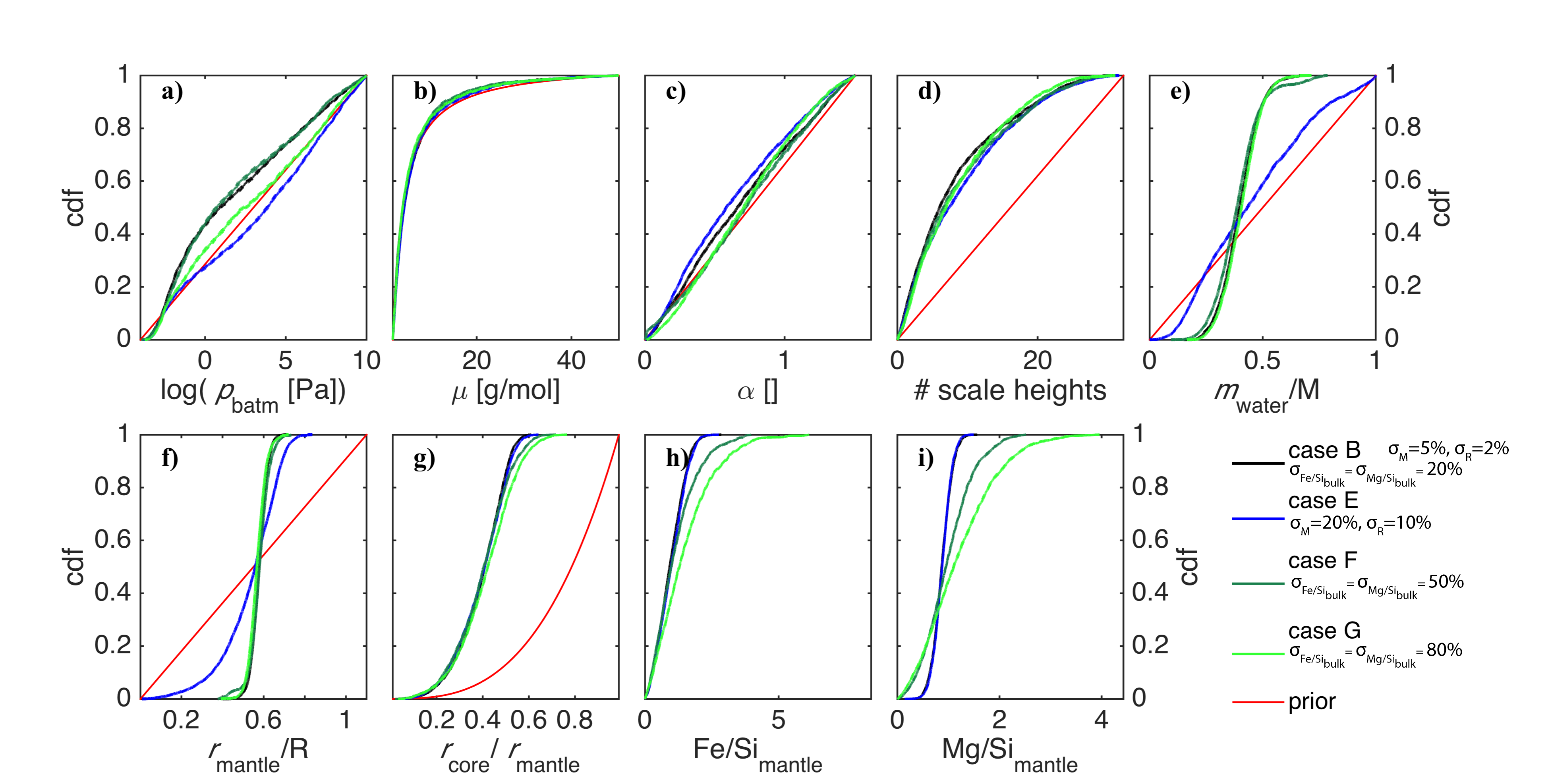}\\
 \caption{Sampled 1-D marginal posterior cdfs of model II parameters for synthetic planet cases B, E, F, G  that vary in terms of data uncertainties. B is the reference case ($\sigma_M$=0.05 M, $\sigma_R$ = 0.02~$R$, 20~\% for both $\sigma_{\fesi}$ and $\sigma_{\mgsi}$), E has larger uncertainties in mass and radius ($\sigma_M$=0.2~$M$, $\sigma_R$ = 0.1~$R$), whereas F and G have larger uncertainties in the abundance constraints, 50~\% and 80~\%, respectively.
(a) Pressure at bottom of atmosphere \Pbatm, (b) atmospheric mean molecular weight $\mu$, (c) temperature-related parameter $\alpha$, (d) number of scale-heights of opaque layers $N$, (e) mass of water \mice, (f) mantle radius \rsolid, (g) core radius \rc, (h) $\fesima$, (i) $\mgsima$.  The priors in (h) and (i) are not shown as not to overload the plot, because they differ among the cases. \label{VaryScale}}
\end{figure*}

 \subsubsection{Influence of data uncertainty}
 
Here, we study the influence of data uncertainty on structural parameter estimation.  As summarized in Table \ref{tabledata}, we vary uncertainty in mass and radius  between cases B ($\sigma_{M}$ of 5\%, $\sigma_{R}$ of 2~\%) and E ($\sigma_{M}$ of 20~\%, $\sigma_{R}$ of 10~\%); we vary uncertainties on planet bulk abundances between cases B (20~\%), F (50~\%), and G (80~\%). All cases B, E, F, and G have the same bulk density of 3.62~g/cm$^3$. The smallest chosen data uncertainties reflect those of high quality data {\coco similar to those expected from PLATO}. Results are shown in Figures \ref{VaryGas} and \ref{VaryScale}. {\coco The results can be summarized as follows:}
 \begin{itemize}
\item Mass and radius uncertainties mainly affect estimates of \rsolid, \mice, and \Zenv. For example, the retrieved confidence region for \rsolid and \mice is three times larger  in case E compared to case B (the 5~\% to 95~\% percentile range of \rsolid for case E is 0.28--0.73~$R$ compared to 0.54--0.66~$R$ in case B; similarly the range of \mice for case E is 0.08--0.93~$M$ compared to 0.22--0.5~$M$ in case B). 
\item Mass and radius uncertainties do not significantly affect estimates of core and mantle composition, since they are conditioned to the same abundance constraints (cf. case B and E).
\item Reducing the uncertainties on the abundance constraints mainly improves the ability to constrain the mantle composition. For example, the 5~\% to 95~\% percentile ranges for $\mgsima$ in cases F and G are larger by a factor of 2.6 and 3.4 compared to case B, respectively. 
\item Compared to the studied cases, the influence on determining core size is more pronounced for purely rocky planets as described by \citet{dorn}. Here, only moderate effects are seen for core size estimates, where the 5~\% to 95~\% percentile range of core size \rc is 30~\% larger for case G compared to B. 
\item Uncertainties on the abundance constraints have only minor effects on estimates of \rsolid and \mice. Between cases B and G, for example, the 50th percentile of \mice varies by up to 8~\%. 
\end{itemize}

For the studied cases, mass and radius uncertainties are more important than uncertainties on $\fesi$ and $\mgsi$ to constrain key structural parameters such as  \mice and \rsolid. This conclusion might vary depending on the actual planet mass and bulk density.

\begin{figure}[ht]
\centering
 \includegraphics[width = 0.48\textwidth, trim = 2cm 0cm 16cm 0cm, clip]{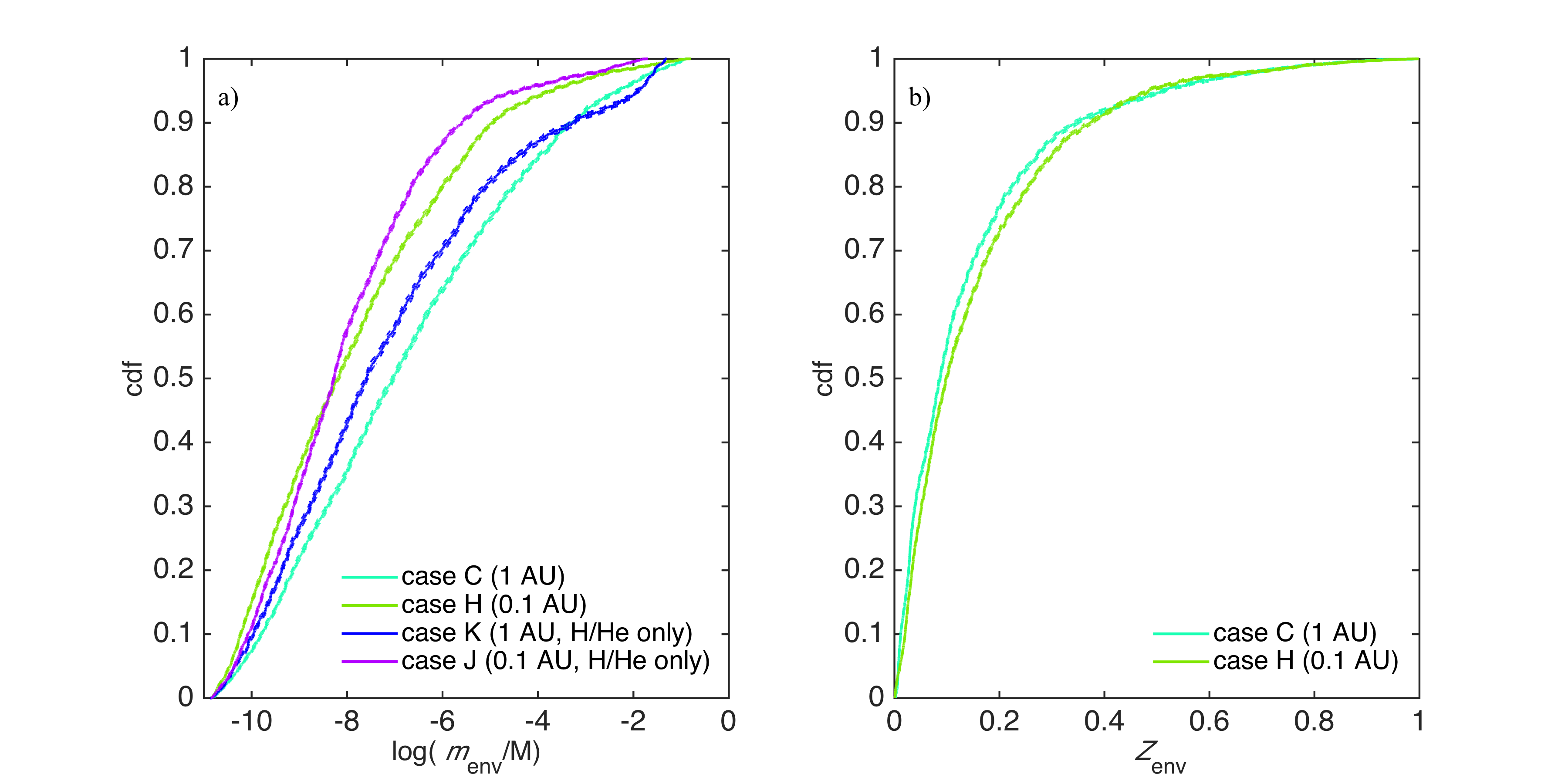}\\
 \caption{Sampled 1-D marginal posterior cdfs of \menv (model I)  for the synthetic planets: case C at 1 AU, case H at 0.1 AU, case J at 1 AU, and case K at 0.1 AU. For cases J and K, the gas composition is restricted to pure H/He (\Zenv = 0) using the EoS of \citet{saumon}. \label{HHe1}}
\end{figure}

 \subsubsection{Influence of semi-major axes}

The semi-major axis influences the energy budget available in the gas envelope and thereby the radius of the planet. Figure \ref{HHe1} demonstrates the effect of distance to the star on estimates of \menv. For the same planet {\coco with a} smaller semi-major axis (case H compared to C), the interior can be explained by a smaller \menv and higher envelope metallicity \Zenv, although the effect on \Zenv is small (not shown). This result is intuitive, since a hotter gas envelope implies a lower gas density, which results in a larger radius. Thus, in order to compensate for a higher intrinsic luminosity while still fitting the radius, the gas mass must be smaller and/or more heavier elements need to be present.
If only pure H/He gas layers are considered, the same trend for \menv is observed (cases K and J in Figure \ref{HHe1}a). Compared to metal-rich envelopes, the restriction to pure H/He envelopes leads to smaller \menv for the reason just discussed.

\subsection{Influence of prior distribution}
\label{prima}
\begin{figure}[ht]
\centering
 \includegraphics[width = 0.5\textwidth, trim = 1cm 0cm 1cm 0cm, clip]{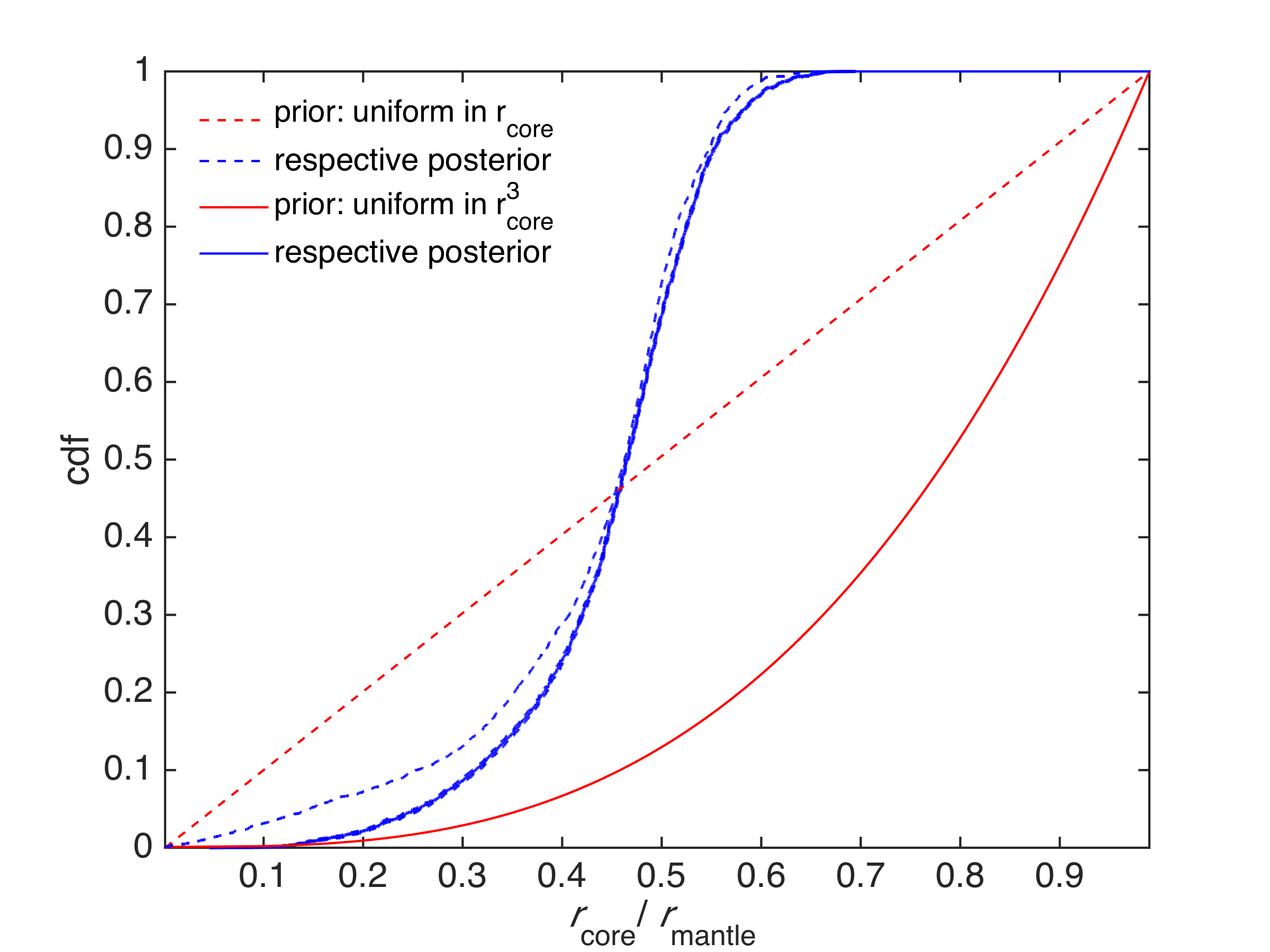}\\
 \caption{\coco Sampled 1-D marginal posterior cdfs (blue) for different priors (red) of core size \rc for Neptune (applying model I). Distributions are depicted in dashed when the prior is uniform in \rc and solid when it is uniform in \rc$^{3}$. The latter is identical to Figure \ref{NepGas}e.)
  \label{priorfigrc}}
\end{figure}

\begin{figure}[ht]
\centering
 \includegraphics[width = 0.45\textwidth, trim = 0cm 0cm 1cm 0cm, clip]{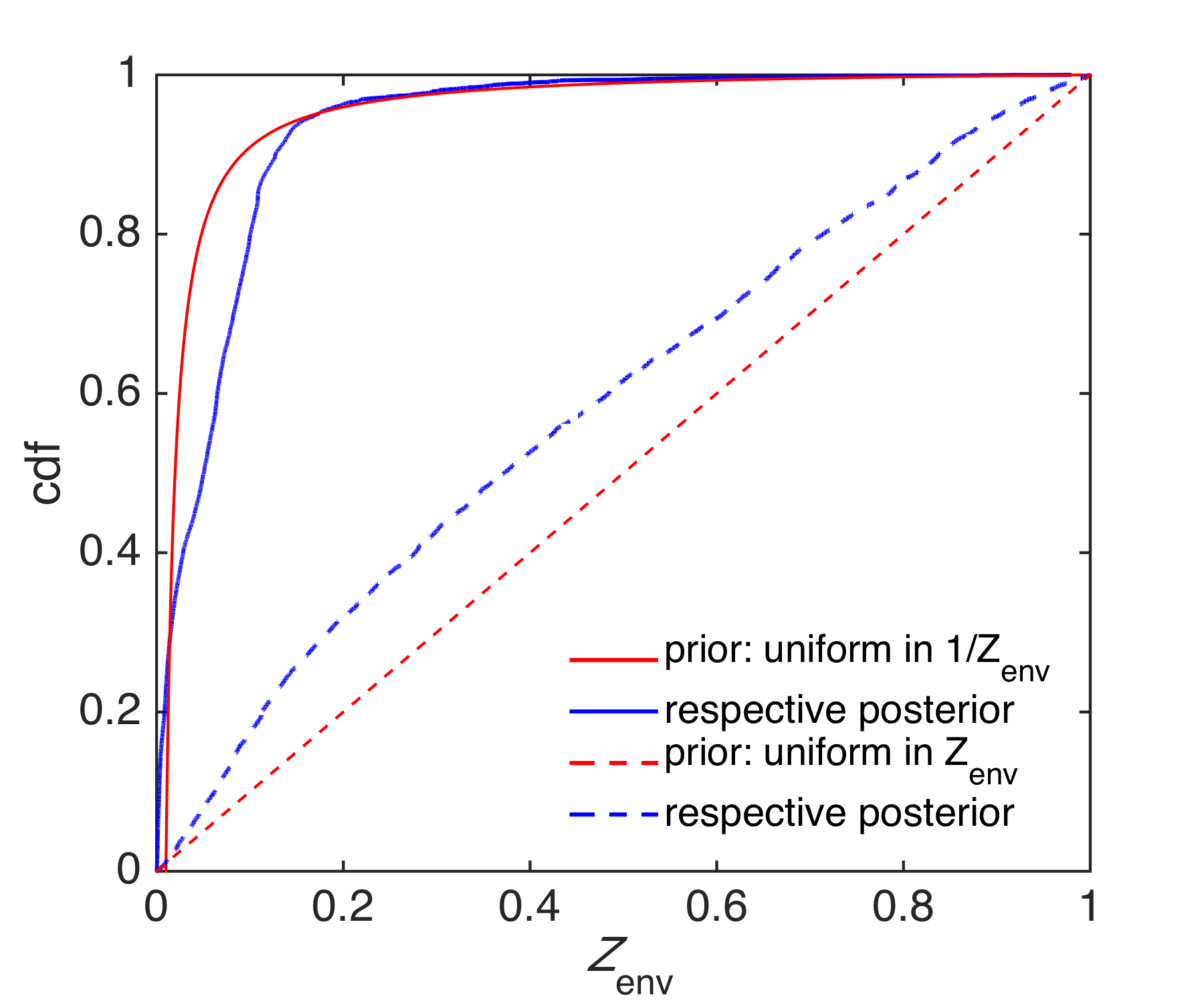}\\
 \caption{\coco Sampled 1-D marginal posterior cdfs (blue) for different priors (red) of envelope metallicity \Zenv for case C (7\ME, 2.6 \RE, applying model I). Distributions are depicted in dashed when the prior is uniform in \Zenv and solid when it is uniform in 1/\Zenv. The latter is identical to Figure \ref{VaryRadii1}c.)
  \label{priorfigz}}
\end{figure}

\begin{figure}[ht]
\centering
 \includegraphics[width = 0.5\textwidth, trim = 0cm 0cm 1cm 0cm, clip]{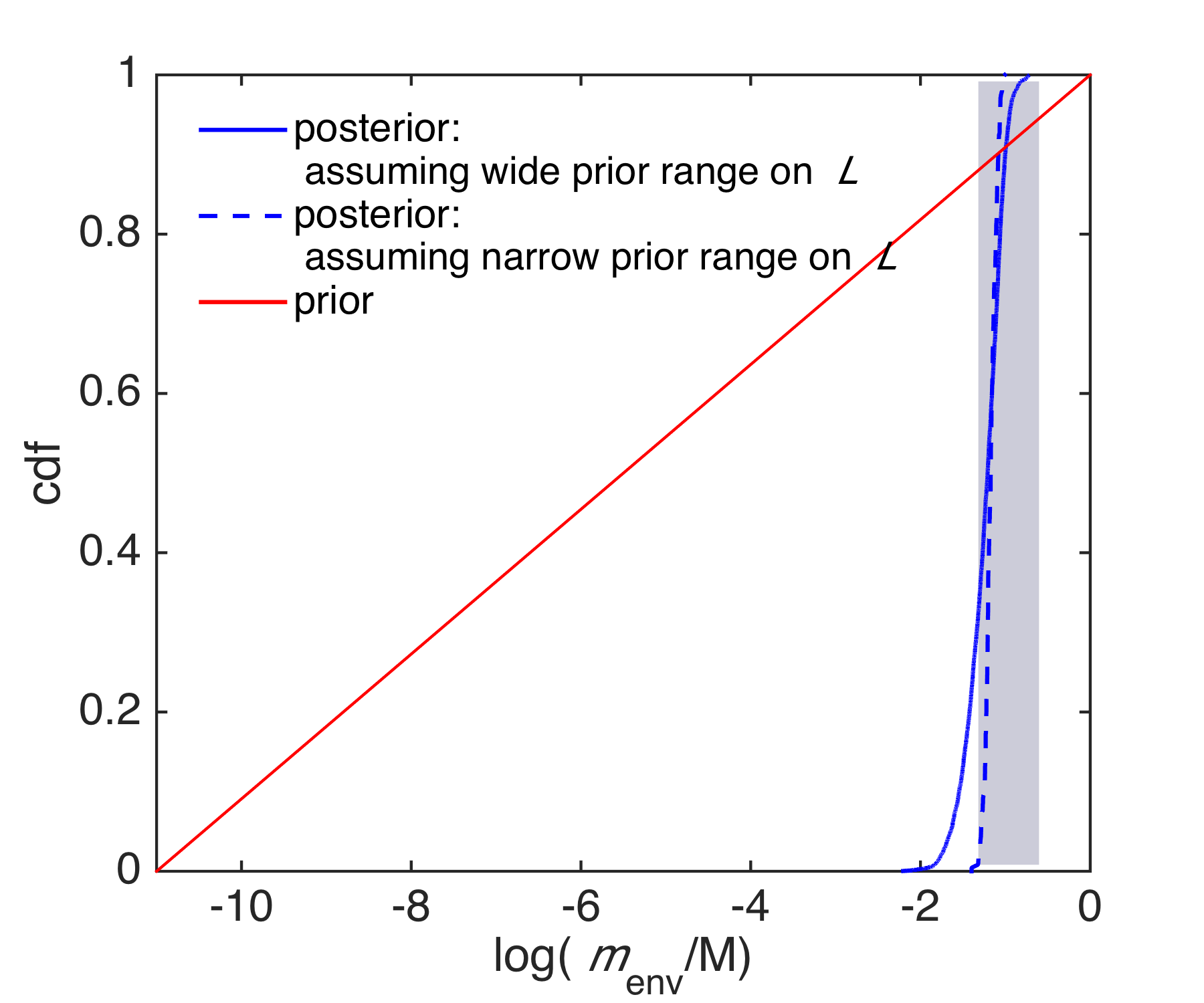}\\
 \caption{\coco Sampled 1-D marginal posterior cdfs (blue) for \mice assuming different priors on $L$ for Neptune (applying model I). Solid blue line refer to wide prior range on $L$ ($10^{18} - 10^{23}$ erg/s), whereas dashed blue line refer to narrow prior range on $L$ ($10^{22.47} - 10^{22.57}$ erg/s). The former is identical to Figure \ref{NepGas}a. Gray area represent independent literature estimates (see main text).
  \label{priorfigl}}
\end{figure}

{\coco The results obtained by a Bayesian inference analysis are subject to the choice of prior, which,
if not chosen carefully can lead to a significant imprint on parameters that are weakly constrained by data. 
In the following, we consider a number of different priors to illustrate this on a selected set of parameters
that are sensed differently by the data considered here. We have singled out core size, which is largely 
determined by bulk abundances and mass, in addition to envelope 
metallicity and luminosity that are mainly constrained by radius and stellar irradiation.

Figure \ref{priorfigrc} illustrates the effect of different prior choices on estimated 
(posterior) core size \rc for a Neptune-sized planet. Here, we contrast a uniform prior in \rc with a 
uniform prior in \rc$^{3}$. A uniform prior in \rc gives more weight to smaller core 
sizes relative to a uniform prior in \rc$^{3}$. But since \rc$^{3}$ is directly proportional to core mass it represents the more natural choice.  
The results indicate that the effect of the prior is negligible for the 
50~\%-percentile of \rc. 
This is an example where the choice of prior is less significant. 

Next, we investigate an example where the estimated parameter is only weakly constrained by data. This is, 
for example, the case for envelope metallicity \Zenv. We compare a uniform prior in \Zenv and in 1/\Zenv for
a case-C planet. A uniform prior in 1/\Zenv is motivated by the fact that H and He are most abundant elements and that 
primary atmospheres are likely rich in H and He \citep[e.g.,][]{alibert}. Also, the scale height of the gas layer 
correlates positively with 1/\Zenv.  
The results are shown in Figure \ref{priorfigz} and illustrate that a uniform distribution 
in \Zenv, relative to a uniform in 1/\Zenv, gives more weight to larger envelope metallicities. This implies
that we are favoring lighter-element atmospheres over heavier-elements. A uniform prior in \Zenv may be more 
appropriate for secondary (outgassed) atmospheres, for which heavy element enrichment is {\textit a priori}
a more likely scenario.

Finally, we consider luminosity $L$. 
For purposes of illustration, we chose the following range 
$10^{22.52\pm0.05}$ erg/s, which corresponds to the observed luminosity of Neptune. More generally,  additional constraints such as infrared flux measurements would allow for 
a narrower prior range on luminosity.  Figure \ref{priorfigl} illustrates the effect of assuming different prior 
ranges on $L$ in estimating gas mass fraction \menv/M for the case of a Neptune-sized planet.  The new prior range on $L$ leads to an improved constraint on gas mass fraction of 
0.05$<$\menv/$M<$0.09 that better predicts independent geophysical estimates relative to the earlier determined range (0.01$<$\menv/$M<$0.2), where a relative wide
prior range was invoked (Table~\ref{tableprior}). In this example, the choice 
of prior has no significant effect on the 50~\%-percentile of \menv/M.

From the above, we can conclude that the posterior distribution is mostly affected by the assumed prior 
distribution for those parameters that are weakly constrained by data. In summary, it should be emphasized that the choice of prior is not arbitrary but need to be based (whenever possible) on observations, laboratory measurements and/or theoretical considerations.
}

\section{Discussion}
\label{Discussion}

Here, we have extended the method of \citet{dorn} from purely rocky exoplanets to general exoplanet types that include volatile-rich layers in the form of water ice, oceans, and atmospheres. For the same data of mass, radius, and bulk abundance constraints, the degeneracy of core and mantle parameters is generally larger in planets of general structure than for purely rocky planets, since their contribution to mass and radius can in part be compensated by volatile material. 

{\coco The key to constrain the structural parameters resides in the large density} contrasts between rock, water, and gaseous layers. In other words, our ability to constrain interiors is because of the different data sensitivity of the various parameters.
The abundance constraints couple core size with mantle size and composition. The relative sizes of core and mantle are thus determined by $\fesi$. The mass of the planet mainly dictates the absolute size of the rocky part and the mass of water. Planetary radius meanwhile determines the characteristics of the envelope and the water layer.

The strength of the presented inference method is that it is modular, i.e., different interior structure models can be tested against each other. However, the applicability and informative value of the inference method is subject to imposed assumptions on the structure model. For example, the two tested atmospheric models differ in terms of complexity and general applicability. 

{\coco Model I is more elaborate in that it calculates pressure-temperature profiles for a given composition while solving for hydrostatic equilibrium, mass conservation, and energy transport.} But it is restricted to H, He, C, and O and it assumes equilibrium chemistry, ideal gas behavior, as well as prescribed opacities. The latter are fit to results of radiative equilibrium models that use a wavelength-dependent opacity function by \citet{JIN2014} for solar metallicities. In that regard, the opacities used are not self-consistent when non-solar metallicities are considered (\Zenv $\ne$ 0.02). Different values of opacities can lead to differences in radius by up to 5~\%. Models that compute line-by-line opacities with their corresponding atmospheric abundances should be performed in the future to compute planetary radii in a self-consistent way. The assumption of ideal gas behavior introduces a bias in radius for large atmospheric mass fractions, for example for a 1~\% \menv/$M$ planet atmosphere the difference in the radius between ideal gas and the \citet{saumon} EoS (for H-He) can reach \mbox{10~\%}.

   Model II assumes an isothermal, homogeneous atmosphere and ideal gas behavior. Therefore, model II is strictly only valid in the case of thin atmospheres (\menv$ \lesssim 10^{-3}$~\ME). While, future available spectroscopic measurements will allow to constrain the key characteristics of the atmosphere \citep{benneke}, it will be difficult to make use of these additional constraints when using the simplified atmospheric model II since isothermal temperatures are non-physical.
However, in the case of thin atmospheres, model II has the advantage of being computationally inexpensive and very general  in the way it is set up, i.e., it does not make assumptions about opacities but fully decouples  structure and opacity of the atmosphere by distinguishing between $\mu$ and $N$, where $N$ accounts for the effect of trace elements in the atmosphere that can have a big impact on opacity.  Therefore, model II is especially useful for secondary atmospheres on small exoplanets, where the composition of the atmosphere can be very diverse. In comparison, model I uses prescribed opacities and thus neglects trace elements. Although not warranted here, it is possible to treat opacities in model I as free parameters to account for trace elements at the cost of increasing the number of parameters.

A further limitation of the structural model is the assumption of a pure iron core. If volatile elements in the core are negligible, this assumption leads to a systematic overestimation of core density and thus an underestimation of core size. In addition, we assume sub-solidus conditions in the rocky interior and a perfectly known EoS for all considered materials. Pressures and temperatures in the various planet cases considered here exceed the ranges that can be measured in the laboratory and while ab initio calculations could fill the gaps, these are not always available. Available EoS include some (mostly unquantifiable) uncertainty \citep[see][for detailed examples]{KhanConnolly2016}.

Here, we have used water as a proxy for the composition of the ice and ocean layers, but other compositions are also possible (e.g., CO, CO$_2$, CH$_4$, NH$_3$). Water is often used as a proxy for ice, since (1) oxygen is more abundant than carbon and nitrogen in the universe, and (2) water condenses at higher temperatures than ammonia and methane.

\section{Conclusions and outlook}
\label{Conclusions}

{\coco We present a} generalized inference method that enables us to make meaningful statements about the interior structure of observed exoplanets. Our full probabilistic Bayesian inference analysis formally accounts for data and model uncertainties, as well as model degeneracy. By employing a Markov chain Monte Carlo technique, we quantify the state of knowledge that can be obtained on composition and thickness of core, mantle, water ice, and gaseous layers for given data of mass, radius, and bulk abundance proxies for $\fesi$ and $\mgsi$ obtained from spectroscopic measurements. 
 We have built upon the work of \citet{dorn} and extended the dimensionality of the interior characterization problem to include volatile elements in the form of gas, water ice and ocean. {\coco Our method succeeds at constraining planet interior structure even for high dimensional parameter spaces and thereby overcomes limitations of previous works on mass-radius relationship of exoplanets.}

We have validated our method against Neptune. Using synthetic planets, we have determined how predictions on interior structure depend on various parameters: bulk density, data uncertainties, semi-major axes, atmospheric composition (i.e., a priori assumption of enriched envelopes versus pure H/He envelopes), {\coco and prior distributions}. Furthermore, we have investigated two different atmosphere models and quantify how parameter estimates depend on the choice of the atmosphere model.
 We summarize our findings as follows:
\begin{itemize}\itemsep0pt
\item It is possible to constrain core size, mantle size and composition, mass of water ice, and key characteristics of the gas layer (e.g., internal energy, mass, composition), given observations of mass, radius, and bulk abundance proxies $\fesi$ and $\mgsi$ taken from the host star.

\item  A Bayesian analysis is key in order to rigorously analyse planetary interiors, as it formally accounts for data and model uncertainty, as well as the inherent degeneracy of the problem addressed here. The range of possible interior structures is large even for small data uncertainties. Our method is able to quantify the probability that a planet is rocky and/or volatile-rich.

\item Our method has been successfully validated against Neptune for which independent structure estimates  based on geophysical data (e.g., gravitational and magnetic moments) are available.

\item Model parameters {\coco have different sensitivity to the various data}. Constraints on bulk abundances $\fesi$ and $\mgsi$ determine relative core size and mantle composition. Mass mostly determines the size of the rocky and icy interior, whereas radius mainly determines structure and composition of the gas and the water ice layers. 

\item Increasing precision in mass and radius leads to a much better constrained ice mass fraction, size of rocky interior {\coco (confidence regions of \mice and \rsolid in case B are three times smaller compared to case E)}, and some improvement on the composition of the gas layer, whereas an increase in precision of stellar refractory abundances enables improved constraints on mantle composition and relative core size.

\item  We have proposed two different atmospheric models: model I solves for radiative transfer; whereas model II uses a simplified scale-height pressure model. Both models yield different insights about possible gas layer characteristics that are subject to prescribed assumptions. In particular, for thick atmospheres, we see a clear discrepancy between model I and II which result in different estimates of  rock and ice layers. The validity of model II is strictly limited to thin atmospheres (\menv $\lesssim 10^{-3}$~\ME). 

\item {\coco We have investigated the effect of prior distribution on estimated parameters and observed that the assumed prior distribution significantly affects the posterior distribution of those parameters,  that are weakly constrained.}

\end{itemize}
In a companion paper \citep{dornA}, we present the application of our method to six observed exoplanets, for which mass, radius, and stellar abundance constraints are available.  

The method presented here is valuable for the interpretation of future data from space missions (TESS, CHEOPS, and PLATO) that aim at characterizing exoplanets through precise measurements of $R$ and $M$. Improving measurement precision, however, is costly as it depends on observation time. Our method helps to quantify the scientific return that could be gained as data precision is increased. Moreover, our study is relevant for the understanding on how interior types are distributed among stars and the implications of these for planet formation.


\begin{acknowledgements}

This work was supported by the Swiss National Foundation under grant 15-144, the ERC grant 239605. It was in part carried out within the frame of the National Centre for Competence in Research PlanetS. We would like to thank James Connolly for informed discussions. 

\end{acknowledgements}

\appendix

\section{Approximation of $\alpha_{\rm max}$}
\label{tlimit}
There is a physical upper limit to the amount of warming by greenhouse gases. The Komabayashi-Ingersoll (KI) limit describes the maximum amount of radiation which can be transferred by a moist atmosphere, which occurs when the transparency $\tau_s$ of the atmosphere becomes very small, i.e., $\tau_s = \tau_{\rm limit}$. 

For model II, this limit is represented by $\alpha_{\rm max}$ and that we roughly approximate as follows:

\begin{equation}
\alpha_{\rm max} = \nicefrac{T_{\rm limit} }{ T_{\rm star} \sqrt{\frac{R_{\rm star}}{2 a}}},
\end{equation}
where $R_{\rm star}$ and $T_{\rm star}$ are radius and effective temperature of the host star, $a$ is semi-major axes, and $T_{\rm limit}$ is:

\begin{equation}
\label{tlim}
T_{\rm limit} = \frac{ T_0}{ln(\frac{\kappa*p_0}{\tau_{\rm limit}g})}.
\end{equation}
Here, $T_0$ is the temperature at some vapor pressure $p_0$ (here, we use $p_0 = 1\times10^{5}$Pa and $T_0 = 373$ K for water, \citep{gold}); $\kappa$ and $\tau_{\rm limit}$ are opacity and optical depth at the KI limit, $g$ is surface gravity. The fraction $\nicefrac{\kappa}{\tau_{\rm limit}}$ is approximated for Earth ($T_{\rm limit} \approx 400$ K) and is estimated to be $10^{-7}$ (in SI units). Thereby, $T_{\rm limit}$ (Eq. \ref{tlim}) scales with the surface gravity. This is a rough estimate for $T_{\rm limit}$ and thus $\alpha_{\rm max}$. More advanced modeling would be required to better determine this limit, but this is outside of the scope of this study.

Equation \ref{tlim} is derived from $\tau_s = \nicefrac{\kappa*p_s}{g}$ and the Clausius-Clapeyron equation, that relates the surface pressure $p_s$ and temperature $T_s$:

\begin{equation}
p_s = p_0 \exp(-\frac{T_0}{T_s}).
\end{equation}



\label{lastpage}

\end{document}